\documentclass[acmsmall]{acmart}

\definecolor[named]{ACMBlue}{cmyk}{1,0.1,0,0.1}
\definecolor[named]{ACMYellow}{cmyk}{0,0.16,1,0}
\definecolor[named]{ACMOrange}{cmyk}{0,0.42,1,0.01}
\definecolor[named]{ACMRed}{cmyk}{0,0.90,0.86,0}
\definecolor[named]{ACMLightBlue}{cmyk}{0.49,0.01,0,0}
\definecolor[named]{ACMGreen}{cmyk}{0.20,0,1,0.19}
\definecolor[named]{ACMPurple}{cmyk}{0.55,1,0,0.15}
\definecolor[named]{ACMDarkBlue}{cmyk}{1,0.58,0,0.21}

\newcommand{\etc}{\emph{etc}\xspace}
\newcommand{\ie}{\emph{i.e.}\xspace}

\newcommand{\eg}{\emph{e.g.}\xspace}

\newcommand{\tname}[1]{\textsc{#1}\xspace}
\newcommand{\tool}{\tname{KaPilot}}

\newcommand{\ttpylines}{2470\xspace}
\newcommand{\ttrslines}{1323\xspace}
\newcommand{\stdfunc}{54\xspace}
\newcommand{\tpfunc}{\revise{70}\xspace}
\newcommand{\ttfunc}{\revise{124}\xspace}
\newcommand{\dataa}{\tname{GoldSet}}
\newcommand{\datab}{\tname{ULSet}}

\newcommand{\textttsmall}[1]{\texttt{\small #1}}
\newcommand{\code}[1]{{\textttsmall{#1}}}

\newcommand{\revise}[1]{\textcolor{black}{#1}}
\newcommand{\revisenew}[1]{\textcolor{black}{#1}}

\newcommand{\specgen}{\revise{\textit{SpecGenerate}}\xspace}
\newcommand{\speceval}{\textit{SpecPrecheck}\xspace}
\newcommand{\specverify}{\textit{SpecVerify}\xspace}
\newcommand{\harngen}{\textit{HarnGen}\xspace}
\newcommand{\safetyreq}{\textit{SafetyReq}\xspace}

\newcommand{\funcsWithGt}{54\xspace}
\newcommand{\funcsNoGt}{\revise{70}\xspace}
\newcommand{\GtPassRate}{88.9\%\xspace}
\newcommand{\NGtPassRate}{\revise{71.4}\%\xspace}
\newcommand{\GtSuccRate}{57.4\%\xspace}
\newcommand{\PassRateMore}{14.8\%\xspace} % succ+bad
\newcommand{\SuccRateMore}{25.9\%\xspace}

\newcommand{\Good}{\revise{Good\xspace}}
\newcommand{\good}{\revise{\emph{good}}\xspace}
\newcommand{\bad}{\emph{bad}\xspace}
\newcommand{\failure}{\emph{failure}\xspace}
\newcommand{\pass}{\emph{pass}\xspace}

\usepackage{graphicx}
\usepackage{listings}
\usepackage{float}
\usepackage{multirow}
\usepackage[ruled,vlined,linesnumbered,noend]{algorithm2e}

\usepackage{stmaryrd}
\usepackage{url}
\usepackage{textcomp} 
\usepackage{caption}
\usepackage{alltt} 
\usepackage{xspace}
\usepackage{verbatim}
\usepackage[inline, shortlabels]{enumitem}
\usepackage{xcolor}
\usepackage{hyperref}
\hypersetup{linkcolor=black,citecolor=black,urlcolor=RubineRed}

\usepackage{enumitem}
\usepackage{fancyvrb}
\usepackage{pifont}
\usepackage{amsmath}
\usepackage{graphicx}
\usepackage{subcaption}
\usepackage{framed}

\setlist[itemize]{leftmargin=*}
\setlist[enumerate]{leftmargin=*}
\setlist{nosep}

\usepackage{xcolor}
\hypersetup{colorlinks,
  linkcolor=ACMDarkBlue,
  citecolor=ACMPurple,
  urlcolor=ACMDarkBlue,
  filecolor=ACMDarkBlue}

\input{_minted/default.style.tex}
\DefineVerbatimEnvironment{MintedVerbatim}{Verbatim}{fontsize=\scriptsize}
\DefineVerbatimEnvironment{MintedVerbatimLineno}{Verbatim}{xleftmargin=2em, numbers=left, fontsize=\tiny}

\renewcommand{\footnotesize}{\scriptsize}

\begin{document}

%%
%% The "title" command has an optional parameter,
%% allowing the author to define a "short title" to be used in page headers.
\title{{\tool}: LLM-Assisted Generation of Kani Specifications for Unsafe Rust Verification}

\author{Minghua Wang}
\authornote{Corresponding author.}
\orcid{0000-0002-2270-2076}
\affiliation{%
  \institution{Ant Group}
  \city{Beijing}
  \country{China}}
\email{minghua.wmh@antgroup.com}

\author{Yuxi Ling}
\authornote{Work done during an internship at Ant Group.}
\orcid{0000-0003-0887-7563}
\affiliation{%
  \institution{National University of Singapore}
  \country{Singapore}}
\email{yuxiling@u.nus.edu}

\author{Mingzhi Gao}
\orcid{0009-0002-3983-4021}
\affiliation{%
  \institution{Ant Group}
  \city{Beijing}
  \country{China}}
\email{gaomingzhi.gmz@antgroup.com}

\author{Yuwei Liu}
\orcid{0000-0001-5170-3388}
\affiliation{%
  \institution{Ant Group}
  \city{Hangzhou}
  \state{Zhejiang}
  \country{China}}
\email{lyw458372@antgroup.com}

\author{Lin Huang}
\orcid{0009-0002-5659-1471}
\affiliation{%
  \institution{Ant Group}
  \city{Beijing}
  \country{China}}
\email{linyu.hl@antgroup.com}

\begin{abstract}
Rust’s ownership and type system provide strong memory safety guarantees, but unsafe code still presents memory safety risks. Formal verification is crucial for ensuring memory safety, but writing precise specifications for unsafe Rust is challenging and largely manual. Large language models (LLMs) have shown promise in generating formal specifications but are often code-centric, prone to inheriting implementation flaws, and lack systematic quality assessment.

In this paper, we present \tool, a multi-agent framework for automatically generating specifications to verify unsafe Rust memory safety using Kani. The process begins with lightweight program analysis and proof harness generation. The \safetyreq agent extracts a concise, refined list of safety requirements from the target Rust function’s documentation, which guides the \specgen agent in producing initial specifications that specify memory safety concerns. Then, the specifications are iteratively refined through a generate–precheck–verify loop involving \specgen, \speceval, and \specverify agents, which assess quality and feed errors back. By executing this loop multiple times, \tool generates a set of candidate specifications. Finally, the shuffle-and-implication strategy is applied to systematically determine the best specification from these candidates. 
We evaluated \tool on \stdfunc unsafe Rust functions with ground truth and \tpfunc without. \tool achieved \GtPassRate and \NGtPassRate specification generation success, respectively, with \GtSuccRate of generated specifications equivalent to or stronger than the ground truth. Compared with AutoSpec, \tool produces \PassRateMore more verifiable specifications and \SuccRateMore more equivalent-or-better specifications. 
\end{abstract}

\maketitle

\section{Introduction}

% % P1: What is the problem we want to solve?
% (draft) Recent work has shown that Generative AI is effective in formal verification tasks in both automatic theorem proving (\eg Dafny, Verus) and interactive theorem proving mechanisms (\eg Lean, $F^*$, Rocq). The proving tasks include mathematical proofs, functional correctness proofs, specification generation, and invariant generation. Methods are either human in the loop or fully automatic.
% %
% However, the state of the art of proof agents does not touch the area of safety verification. Safety verification is harder and equally important. In this paper, we extend LLMs' ability to safety verification by proposing a fully automated LLM agent, \tool. (need to connect this paragraph with Rust and Kani)

% With the growth of the Rust community --> more and more important and widely used --> we need to verify the safety of Rust code.
% Unsafe Rust code is XXX and widespread in Rust libraries. It produces XX problems. However, the verification of unsafe Rust code is hard because: 1) lack of fundamental definition, 2) lack of verification mechanism, 3) slow, 4) XXX.
% %
% What have people done in this area? 
% To prove the safety of unsafe Rust code, XXX has proposed XX, ...
% %
% % Verus VS. Kani

% % P2: What is missing? Research gap/ challenges
% % Automation, Scalability, Precision

% LLM-based Rust specification generation presents several challenges.

Rust provides strong memory-safety guarantees at compile time through its strict type system and ownership model. However, Unsafe Rust allows programmers to explicitly bypass certain compiler checks in exchange for greater performance or expressiveness. While indispensable in systems programming, unsafe code introduces the risk of undefined behaviour (UB), which can lead to severe memory-safety vulnerabilities. Formal verification offers a principled solution by reasoning exhaustively about program behaviour under precise assumptions, enabling reliable proofs of the absence of UB. However, formalisation of safety properties, \ie, writing specifications, especially for Unsafe Rust, as the very first step of formal verification, relies heavily on manual effort from human experts, making it time-consuming and error-prone. To make the formal verification scalable, automated specification generation is desired.

% However, formal verification cannot be applied directly to real-world programs without precise function specifications, which define the conditions under which a function is guaranteed to execute safely.

% Specification generation, therefore, plays an important role in verifying Unsafe Rust code. On the one hand, specifications explicitly capture the safety requirements that an unsafe function must satisfy to preserve memory safety, allowing verification tools to check whether UB can occur under these assumptions. On the other hand, once a function is verified to be free of UB, its specification can serve as a sound safety abstraction for verifying upstream callers, avoiding repeated analysis of low-level implementations and significantly improving verification scalability.

Recently, large language models (LLMs) have demonstrated strong capabilities in code-related tasks, motivating LLM-assisted specification generation~\cite{autospec,specgen,autoverus,alphaverus,llm_dafny,safe}. 
For example, AutoSpec~\cite{autospec} generates verifiable ACSL specifications by iteratively interacting with Frama-C, while SpecGen~\cite{specgen} synthesises specifications directly from source code and refines them through mutation strategies when verification fails. 
Although these LLM-assisted approaches lower the barrier to specification generation, they all share the following challenges in the context of Unsafe Rust: 

% that directly affect specification reliability and the effectiveness of formal verification:

\noindent{\textbf{C1}: Code-centric specification generation inherits implementation flaws.}
Existing approaches~\cite{autospec, specgen} derive specifications directly from the code. When the implementation contains defects or fails to explicitly expose its safety assumptions, the resulting specifications tend to inherit these limitations, reducing their effectiveness in uncovering missing or violated safety requirements.

\noindent{\textbf{C2}: LLM-generated specifications lack stability.}
LLMs may produce specifications with inconsistent quality, including incomplete coverage of documented safety concerns, overly restrictive preconditions, overly permissive postconditions, and limited robustness across different instantiations.

\noindent{\textbf{C3}: Specification quality evaluation is hard to automate systematically.}
Prior works~\cite{autospec,specgen,autoverus} rely on manual inspection to evaluate specification quality. Verification success implies syntactic correctness and verifiability, without ensuring alignment with safety requirements. For example, contradictory preconditions may degenerate into vacuous specifications, allowing verification to succeed without meaningful program behaviour constraints. Moreover, the strength or completeness of specifications lacks objective automated evaluation approaches, leading to subjective and potentially inconsistent assessments.
%%%

% TODO 让这一段介绍更简洁，更突出重点
To address these challenges, we propose \tool, a multi-agent framework of automated specification generation to verify unsafe Rust code in Kani. \tool consists of five specialised agents: 
Safety Requirement Analysis (\safetyreq), Harness Generation (\harngen), Specification Generation (\specgen), Specification Precheck (\speceval), and Specification Verification (\specverify), that collaboratively generate, assess, validate, and refine formal specifications.
% Safety Requirement Analysis, Harness Generation, Specification Generation, Specification Precheck, and Specification Verification—that collaboratively generate, assess, and validate formal specifications.

\tool takes the documentation, instead of the code, as the primary source to specify safety properties for the target function. Unlike functional correctness, which can be inferred from the code, hidden safety properties are usually revealed by the documentation. By summarising an unstructured description in natural language to a concise list of safety requirements, \tool produces a more structured target for subsequent specification generation (addressing \textbf{C1}).

% The \safetyreq agent takes the target function’s documentation as its only input and extracts safety concerns using well-designed extraction and evaluation rules. Through iterative generation and assessment, it produces a concise list of safety requirements, as a reliable target for subsequent specification generation (addressing \textbf{C1}).

Guided by this safety requirement list, \specgen constructs formal specifications using Kani primitives.
% generating nondeterministic inputs and invoking the target functions. 
Instead of sending them to Kani for verification, \speceval performs a precheck using LLMs, assessing whether they fully cover all documented safety concerns and whether their semantics are aligned with safety requirements, not overly strong or weak. Specifications that fail this precheck are fed back to the \specgen agent for refinement and re-assessment, ensuring the stable quality of specifications before the verification (addressing \textbf{C2}).

Finally, \tool invokes Kani to verify the specifications against the target function. A generate–evaluate–verify loop is designed to iteratively repair the specification. 
% Verification errors are categorised and fed back for repair. Syntax errors trigger regeneration based on Kani diagnostics, while property violations are addressed using counterexamples and comparisons between failing and successful harnesses. 
Particularly, when verification succeeds, vacuity checks are applied to ensure that postconditions are not trivially satisfied due to contradictory preconditions. After \tool constructs a set of specification candidates, a \emph{shuffle-and-implication} strategy is applied to select optimal preconditions, postconditions, and loop invariants. \tool derives high-quality final specifications (addressing \textbf{C3}).

We evaluate \tool on two benchmark datasets: \funcsWithGt Rust functions with ground-truth specifications (\dataa) and \funcsNoGt functions without ground truth (\datab). On \dataa, \tool achieves an \GtPassRate specification generation success rate, with \GtSuccRate of the generated specifications being semantically equivalent to or better than the ground truth. On \datab, \tool successfully generates specifications for \NGtPassRate of the functions. Compared with AutoSpec on \dataa, \tool achieves a \PassRateMore higher success rate in generating verifiable specifications and a \SuccRateMore higher success rate in generating specifications that are semantically equivalent to or better than the ground truth. Apart from that, ablation studies confirmed the effectiveness of \tool's components.

\paragraph{Contributions.} We summarise the contributions as follows:

\begin{itemize}
    \item We present and open-source \tool\footnote{https://github.com/MinghuaWang/KaPilot}, a fully automated, multi-agent framework for generating Kani specifications to verify memory safety in unsafe Rust code. We design safety requirement extraction that improves LLMs in specifying safety properties. We also introduce a lightweight \emph{shuffle-and-implication strategy} to \tool that further improves the specification quality.

    \item We design a multi-agent specification generation pipeline that decomposes the verification task into specialised agents. These agents collaborate in a generate-precheck-verify loop, enabling specification generation not only verifiable but also semantically aligned with refined safety requirements and of appropriate strength.

    \item We conduct an extensive evaluation of \tool on \dataa and \datab datasets, demonstrating its effectiveness in the safety verification of unsafe Rust code. 

\end{itemize}

\section{Background}
    
\subsection{Unsafe Rust and Undefined Behaviour}

Rust allows unsafe operations that can potentially violate the memory-safety guarantees of the Rust compiler and transfer the responsibility of ensuring the code safety from the Rust compiler to programmers, that is, unsafe Rust~\cite{rustbook}. The \code{unsafe} gives programmers the power to dereference a raw pointer, access or modify a mutable static variable, and access fields of \code{union}S. It is easy to identify unsafe Rust code, since it must be marked with the \code{unsafe} keyword, \eg, \code{unsafe fn}, \code{unsafe trait}, and \code{unsafe \{\}}. \autoref{lst:unsafe_rust_sub1} presents an example excerpted from \code{"library/core/src/ffi/c\_str.rs"} in the Rust standard library. To calculate the length of a null-terminated string, it dereferences the raw pointer \code{ptr} and calls an external function from the C standard library with safety condition statements specified at lines 11 and 19. These safety conditions can be expressed informally, as natural language comments, or formally, as explicit preconditions. The safety of the unsafe Rust code critically depends on these safety conditions being upheld by programmers. 

Rust programmers are encouraged to wrap unsafe code within safe abstractions and expose only safe APIs. Such safe abstractions are pervasive in the Rust standard library, where unsafe implementations cover many core functionalities. Within these abstractions, programmers are responsible for ensuring that the calling of unsafe code satisfies its safety condition in the safe abstraction. Otherwise, the violation of safety conditions may lead to undefined behaviour (UB), thereby bringing the regular Rust potential memory safety problems. The Rust Reference~\cite{ublist} provides a not exhaustive list of UB, like mutating immutable bytes, producing invalid values, \etc.
% In this work, we focus on the safety verification of unsafe Rust, proving the absence of XX undefined behaviour.

\begin{figure}[t]
    \centering
    \begin{subfigure}{0.48\textwidth}
        \centering
        % \inputminted[linenos,xleftmargin=2em,fontsize=\tiny]{rust}{code/unsafe_example.m}
        % \input{code/unsafe_example.m}
        \input{_minted/code/unsafe_example.m}
        \caption{The unsafe Rust function}
        \label{lst:unsafe_rust_sub1}
      \end{subfigure}
      \hfill
      \begin{subfigure}{0.48\textwidth}
        \centering
         % \inputminted[linenos,xleftmargin=2em,fontsize=\tiny]{rust}{code/kani_example.m}
         % \input{code/kani_example.m}
         \input{_minted/code/kani_example.m}
        \caption{The Kani contract and harness}
        \label{lst:unsafe_rust_sub2}
      \end{subfigure}
    \caption{An unsafe function from the Rust standard library and its Kani verification}
    \label{lst:unsafe_rust}
\end{figure}

% std version: bacd51cab7165c83c40b6dcceae0682e4bffd74c.
% a table: solved challenges & functions
\subsection{Safety Verification in Kani}
\label{sec:back_kani}
Kani~\cite{kani} is a bit-precise bounded model checker for Rust that supports the verification of unsafe Rust code. It detects memory-safety violations such as null pointer dereferences and use-after-free, as well as runtime panics caused by behaviour including out-of-bounds accesses and arithmetic overflows. Kani performs bounded model checking using proof harnesses with symbolic inputs, exhaustively exploring program executions up to a given bound.
Recent versions of Kani provide specification primitives for writing function contracts and loop invariants, enabling modular verification. Functions can be verified against their specifications and then treated as safe abstractions at call sites, allowing verification results to be composed across the program. This significantly reduces verification cost for code with deep call chains or repeated function invocations.

To illustrate how Kani verifies the safety of an unsafe Rust function, \autoref{lst:unsafe_rust_sub2} shows the code added with the Kani contract and harness. 
The precondition and postcondition with \code{requires} and \code{ensures} clauses specify the safety requirements given by developers in the function's descriptions. 
%As we can see, it specifies the absence of XXX UB. XXX more explanation. 
The \code{check\_strlen\_contract} harness function at line 23 is the entry point of Kani verification. The annotation \code{kani::proof\_for\_contract} specifies that the Kani will verify the function via contracts at lines 1-20. In the Kani harness, it declares an arbitrary string of length 32 as the input parameter of the \code{strlen} function and invokes the function within an unsafe block. Then, Kani will perform model checking and check whether the specification is satisfiable.

\begin{section}{Approach}
\label{sec:approach}
% care about only memory safety and Rust panics

% 可能挪到introduction
% Unlike general functional correctness verification, where properties to verify can be explicitly inferred from the code, safety requirements are usually hidden and hard to extract, making the safety verification difficult for LLMs. To overcome this challenge, we borrow the workflow that human experts use and design a multi-agent framework for the automated verification of unsafe Rust programs in Kani, named \tool, with a specific LLM agent at each of the key steps.

\subsection{Overview}
\label{sec:overview}

% (listing everything here would make readers confused)\tool consists of a safety requirement analysis agent (\safetyreq), a harness generation agent (\harngen), a specification generation agent (\specgen), a specification evaluation agent (\speceval), and a specification verification agent (\specverify), which together form an iterative generate–evaluate–verify pipeline.
We propose \tool, a multi-agent framework to automate the specification generation for unsafe Rust programs in Kani. ~\autoref{fig:arch} shows the overview of \tool.

It takes the source code of the target functions and their documentation as inputs. In the preparation stage, it first extracts metadata from the source code via lightweight program analysis, including function signatures, call graph, and function description. Then, agent \harngen constructs a Kani proof harness for the target function (details in \autoref{sec:preparation}).
Meanwhile, \safetyreq extracts safety requirements based on the documentation written in natural language (details in \autoref{sec:safetyreq}). For every processed function, its metadata and safety requirements will be stored in a shared knowledge database for further \harngen and \specgen.

\begin{figure}[t]
    \centering
    \includegraphics[width=0.9\linewidth]{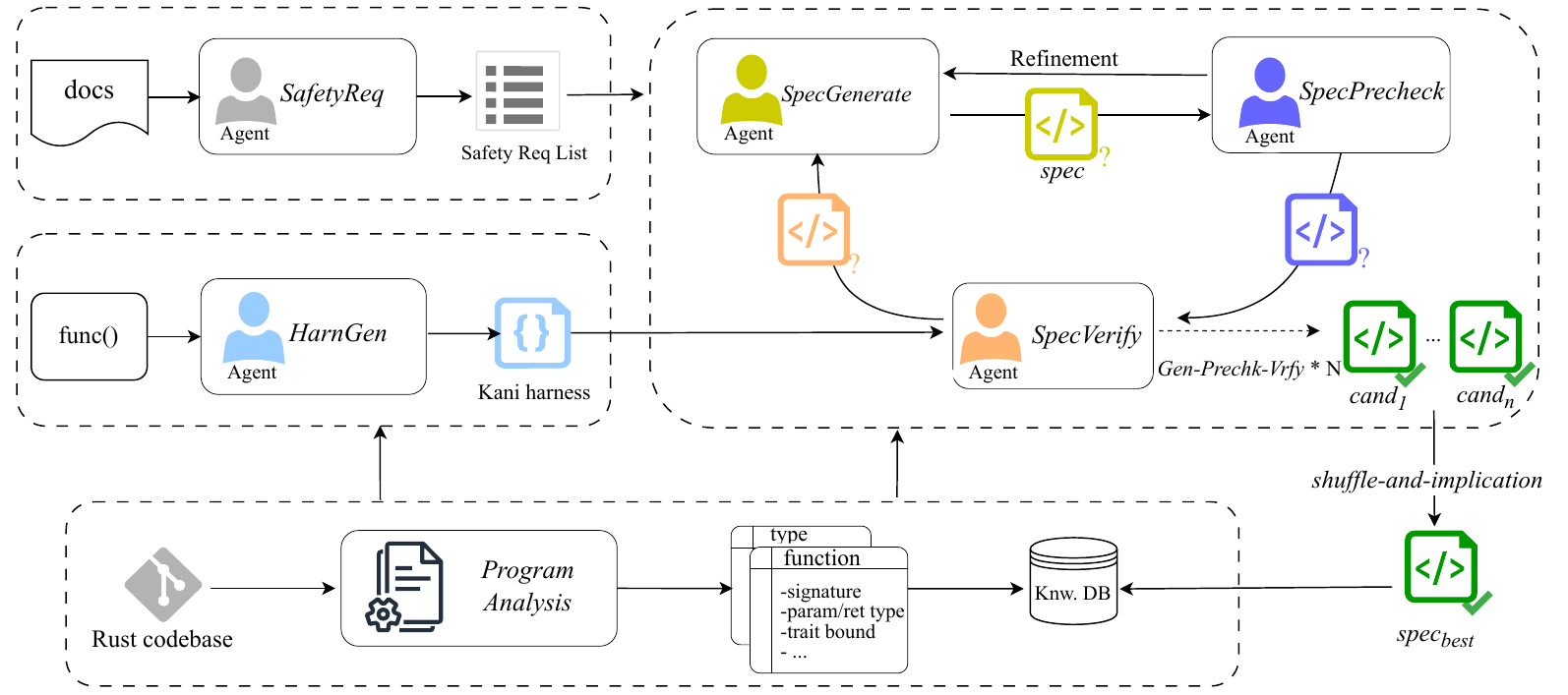}
    \caption{The pipeline of \tool.}
    \label{fig:arch}
\end{figure}

% Given a target function, we first perform lightweight program analysis to extract function signatures, type definitions, and documentation comments and store them in a shared knowledge database. 

%Based on the function documentation, the agent \safetyreq extracts safety-related constraints and summarizes them into a concise set of safety requirements. 

With the Kani harness and safety requirements ready, the agent \specgen generates formal specifications for the target function consisting of preconditions, postconditions, and loop invariants using Kani primitives (details in \autoref{sec:specgen}). 
Before sending it to Kani to verify, \tool performs a precheck in \speceval to eliminate simple syntax errors and check whether it covers all mentioned safety requirements. If the specification gets a low score, it would be fed back to \specgen for refinement. \tool repeats this step till it reaches the maximum iterations or passes the checking of \speceval (details in \autoref{sec:speccheck}). 
%The agent \speceval assigns quantitative scores based on safety requirement coverage and specification strength, and the specification with low scores is fed back to \specgen for regeneration. 
The passed specification is submitted to the \specverify agent, which invokes Kani for bounded model checking. Specifications that successfully pass the verification in Kani are marked as candidate specifications (details in \autoref{sec:specverify}).

Although candidate specifications have been evaluated and verified by the \speceval and \specverify agents, their quality may still be suboptimal, for instance, with overly restrictive preconditions or insufficiently precise postconditions. To mitigate this issue, the \emph{specification shuffle and implication strategy} is designed to select the best specification from the candidates (details in \autoref{sec:specselect}). Once the final specification is determined, \tool stores it in the shared knowledge database. These high-quality specifications can be retrieved and reused as few-shot examples to guide LLMs in generating specifications for new Rust functions.

\subsection{Preparation}
\label{sec:preparation}
% or preprocessing
The preparation stage here performs preprocessing on the existing codebase and documentation, containing two tasks: 1) metadata extraction; 2) the Kani harness generation.

\subsubsection{Metadata extraction.}
\label{sec:metadata}
In this step, \tool extracts the metadata of the target function via a lightweight static program analysis, including function signature, function descriptions, and available public methods in the code base. 
When identifying a Rust type, particularly since it can be either concrete or a trait, we record its definition and, additionally, the inherited traits if it is concrete and concrete types implemented from it if it is a trait.
Additionally, we extract available public methods for each target function, which can serve as auxiliary functions when constructing preconditions, postconditions, and loop invariants in later specification generation. 
All metadata is recorded in a shared knowledge database for later specification generation.

% For Rust functions, we extract parameter types, return types, and documentation comments preceding the function signature. 
%These comments describe the function’s behaviour and relevant safety concerns, serving as the primary source for generating formal specifications. 

% If formal specifications already exist for a function, we also extract and store them in a shared knowledge database. These existing specifications can serve as few-shot examples to guide LLMs when generating specifications for new functions.

\subsubsection{Harness generation.}
\label{sec:harngen}
As mentioned in \autoref{sec:back_kani}, the Kani harness is essentially a Rust function that invokes the target function with all input parameters initialised in Kani types, such as \code{kani::any()}, which serves as the entry point of Kani. The agent \harngen constructs the proof harness for the target function using AutoHarness~\cite{autoharness}. However, it cannot handle complex scenarios: parameters are not Rust primitive types; parameters require extra implementation of traits beforehand; user-defined data types cannot be directly supported in Kani; \etc. To make this step fully automatic, \harngen employs LLMs to fill in necessary but missing parameter declarations for failed cases from AutoHarness and checks target function is invoked correctly in the harness. 

% Following program analysis, the workflow invokes the agent \harngen, which constructs a proof harness for the target function by generating nondeterministic inputs for its parameters and invoking the function within the harness. We use AutoHarness ~\autoref{autoharness} to generate the harness. However, it currently supports only Rust primitive types and cannot handle more complex scenarios. For example, parameters of trait types require generating all concrete types that implement the trait, and user-defined complex types cannot be directly supported. For these cases, we employ HarnessLLM~\autoref{harnessllm} to generate the necessary nondeterministic parameters. The resulting harness establishes the execution environment required for bounded model checking with Kani.

\subsection{Safety Requirements Extraction}
\label{sec:safetyreq}
% The agent \safetyreq takes the documentation of the target function as the input and employs LLMs to distill it into a concise list of atomic safety requirements. \autoref{fig:safety_req_exmaple} presents an example that converts the documentation of \code{offset} function on the left to a fine-grained safety requirement list on the right down to the constraints of specific program variables, \code{count * size\_of::<T>()}. This list will serve as the specification target for subsequent specification generation and verification stages.
Before the specification generation, we need to identify the safety properties to verify for the target function. Unlike functional correctness, safety properties are usually hidden in the code. The most direct source of such properties is the documentation, \ie, function description, which is written in natural language. 
However, such descriptions are often incomplete, distributed across cross-referenced APIs, or encoded implicitly as behavioural constraints rather than explicitly stated in \code{Safety} or \code{Panics} sections (e.g., pointer-distance computations must not overflow \code{isize}). And descriptions sometimes use variables that are not aligned with function parameters, or safety related description are encoded implicitly as behavioural constraints rather than explicitly stated in \code{Safety} or \code{Panics} sections (e.g., pointer-distance computations must not overflow \code{isize}).
For example, in \autoref{fig:doc-offset-from-unaligned}, the documentation of \code{offset\_from\_unsigned} does not fully contain all relevant safety concerns, and highlighted lines require cross-referencing the related function \code{offset\_from} elsewhere in the codebase, as shown in \autoref{fig:doc-cross-ref-offset-from}.

In this case, we design an agentic LLM, \safetyreq, that iteratively extracts, evaluates, and refines structured safety requirement lists from unstructured and possibly low-quality documentation, which serves as the specification target for subsequent specification generation and verification stages. \autoref{fig:safety-req-extracted} presents the resulting safety requirement list distilled by \safetyreq.

\begin{figure}[t]
    \centering
    \captionsetup[subfigure]{font=scriptsize}
    \begin{subfigure}[t]{0.46\textwidth}
        \input{_minted/tex/desc-offset_from_unsigned}
        \caption{\code{offset\_from\_unsigned}'s safety concerns, some of which are documented in \code{offset\_from}.}
        \label{fig:doc-offset-from-unaligned}
    \end{subfigure}
    \hfill
     \begin{subfigure}[t]{0.51\textwidth}
        \input{_minted/tex/desc-offset_from.tex}
        \caption{\code{offset\_from}'s safety concerns.}
        \label{fig:doc-cross-ref-offset-from}
    \end{subfigure}
    \par\vspace{2ex}
    \begin{subfigure}[b]{0.99\textwidth}
        \centering
        \input{_minted/tex/safetyreq-offset_from_unsigned.tex}
        \caption{The safety requirements for \code{offset\_from\_unsigned}}
        \label{fig:safety-req-extracted}
    \end{subfigure}
    \caption{An example of safety requirements extraction in \safetyreq.}
    \label{fig:safety_req_exmaple}
\end{figure}

First, we identify the allowed sources from which safety requirements may be extracted: the \textit{Safety} and \textit{Panics} sections in the target function’s preceding documentation, as well as relevant safety-related descriptions introduced through cross-references. Then, the LLM is constrained to generate safety requirements exclusively from these sources and to follow the explicit rules below:

\begin{framed}
\begin{itemize}
\item (R1) Scope. Each requirement must constrain a property related to memory safety or Rust runtime panics of the target function, or the function behaviours related to runtime panics.
\item (R2) Atomicity. Each requirement must consist of a single sentence describing exactly one safety property in a precise and concise manner.
\item (R3) Variable Reference: A requirement may refer only to the target function’s parameters, return values, or generics; if derived from a cross-reference, all referenced variables must be correctly mapped to the target function’s interface.
\item (R4) Source Traceability: Each requirement must end with a citation in the form <src>-<sec>: <statement>, where <src> denotes either the target function (TF) or a cross reference (CR), <sec> denotes a \textit{Safety} or \textit{Panics} section; <statement> is the original documentation text.
\item (R5) Exclusivity. There must be no duplicates between requirements, no logical implications between two requirements, and no contradictions between two requirements.
\item (R6) Completeness. They cover all safety constraints from the allowed sources.
\end{itemize}
\end{framed}
Rules R1–R3 ensure that each extracted safety requirement precisely constrains the memory safety behaviour of the target function, while R4 enables subsequent agents to validate the provenance of each requirement. In addition, rules (R5) Exclusivity and (R6) Completeness ensure that different safety requirements describe distinct safety concerns and that the entire set collectively covers all safety-related statements in the documentation. 
%
% A complete list of rules and corresponding prompts in this step can be found in Appendix~\autoref{xxx}.

After a safety requirement list is generated, the agentic LLM scores it by calculating the harmonic mean of satisfied rules over all rules and plans the next step based on the score. 
% The score aggregates the performance across all rules. 
If the score exceeds a predefined threshold, the current requirement list will be accepted as the final result. Otherwise, the per-rule scores will be fed back to the LLM to guide targeted refinement. If the number of iterations exceeds a predefined limit, the agent will return the highest-scoring safety requirement list observed during the process. As for the experiments in \autoref{sec:rq1}, the threshold score is set to 6, with a maximum of 3 iterations.
%
%

% \begin{equation}%\tag{Formula \theequation}
% \label{eq:safety_req_score}
% \scriptsize
%     score=\alpha \cdot\frac{1}{n} \sum_{i=1}^{n} s_i + \beta \cdot \frac{n}{\sum_{i=1}^{n} \frac{1}{s_i}},  \alpha + \beta = 1
% \end{equation}

% TODO: add an example.

\subsection{Specification Generation}
\label{sec:specgen}
We design the \specgen agent that generates the specification for the target function based on the following sources: 1) primarily the safety requirement list from \safetyreq; 2) the metadata of the target function; 3) Kani's domain-specific knowledge, including the primitives and APIs for expressing preconditions, postconditions, loop invariants, \etc; 4) few-shot examples, if any, from a shared knowledge database; 5) feedback information from the last generation iteration, if any. 

The safety requirement list and metadata of the target function serve as a structured and precise target for the specification generation in this stage. Each requirement captures an atomic memory-safety or runtime-correctness property of the target function and is explicitly grounded in the function documentation. The \specgen agent required that the generated formal specifications must be aligned with all safety requirements.

Note that we exclude the source code of the target function, because we believe that the actual implementation would interfere with LLMs' understanding of safety properties. Intuitively, the source code does not provide useful information to LLMs on how to formally specify safety properties, while it is helpful for functional correctness verification. Furthermore, LLMs tend to follow the source code to specify safety properties if it exists, in a way that would always lead to wrong results. Thus, we replace the source code with the metadata for LLM in generating specifications. Our experiments also evidence this observation (will be detailed in \autoref{sec:rq1}).

% The metadata of the target function extracted in \autoref{sec:metadata} is also necessary for LLM to 

%Moreover, precise specifications often require reasoning about function parameters and return values, as well as invoking methods defined on their data types to capture richer semantic constraints. Without explicit knowledge of these available methods and their functionalities, LLMs are limited in their ability to express accurate specifications. To address this limitation, we provide the LLM with the methods implemented by the parameter and return value types of the target function, together with their documented descriptions. This allows the LLM to retrieve and invoke suitable methods when constructing specifications, leading to more expressive and semantically accurate constraints.

% avoiding both overly restrictive constraints that unnecessarily limit the input space and overly weak constraints that fail to ensure memory safety or prevent runtime panics.

The remaining three sources are intended to further improve the performance of \specgen.
Kani's domain-specific knowledge is injected into LLMs via the shared knowledge database because it helps the LLM select and compose appropriate primitives correctly when constructing specifications. Specifically, this knowledge is mainly Kani specification primitives used in encoding safety properties, including the usage of APIs for expressing pre- and post-conditions, loop invariants, and memory-safety related predicates, such as \code{\#[requires()]} and \code{\#[ensures()]}.

%The specifications use Kani primitives to encode safety properties. To improve the success rate and accuracy of specification generation, we inject domain-specific knowledge about Kani into LLMs. This knowledge summarizes Kani specification primitives, including the signatures and semantics of APIs for expressing preconditions and postconditions, defining loop invariants, and specifying memory-safety predicates. Providing this Kani-specific knowledge helps the LLM correctly select and compose appropriate primitives when constructing specifications.

In addition, we adopt a few-shot \revise{in-context} learning strategy in \specgen. By default, all specifications that are successfully verified in the loop, together with their target functions, will be stored in the shared knowledge database. When \specgen has a new target function, it retrieves the top N functions that are most similar in terms of function code and documentation, then constructs pairs of function code and corresponding specifications as few-shot examples. These few-shot examples offer concrete references on writing Kani specifications, thereby improving the quality of the generated specifications further.

\subsection{Specification Precheck}
\label{sec:speccheck}
In this step, we design \speceval agent to perform a precheck on the quality of the specification from \specgen before sending it to Kani for verification.

%To mitigate these issues, \speceval systematically evaluates the Kani specifications produced by \specgen before they are submitted to formal verification. 
Specifically, \speceval evaluates the generated specifications in two steps. First, it checks whether all safety requirements in the safety requirement list are covered. All uncovered requirements will be fed back to \specgen, which is then instructed to generate specifications targeting the missing requirements. This checking helps solve the side effect of LLM's hallucination and nondeterminism from \specgen.
Second, \speceval asks LLM to assess the strength of the generated specifications by assigning a quantitative score ranging from 1 to 10. It evaluates whether preconditions are overly restrictive with respect to the corresponding safety requirements, and whether postconditions are overly weak and thus insufficient to characterise the function’s safety properties. If the assigned score falls below a predefined threshold, \speceval produces concrete refinement suggestions and feeds them back to \specgen for the refinement.
%
% 单个一条spec -- safety requirement问LLM，语意符合，1-10， 6分至少。5次，超了-》历史最好。

The refinement iteration between \specgen and \speceval continues until all safety requirements are covered and the evaluation score is higher than the threshold, or it reaches the maximum number of iterations, at which point the highest-scoring specification is returned. In our experiment, we set the threshold score to 6 and the iteration limit to 5. 

Note that \speceval is designed as a lightweight semantic-based specification checker rather than a sound verifier, pruning low-quality specifications before verifying them in Kani. Our ablation study (see \autoref{sec:rq_ablation}) evidences that this refinement process helps ensure the resulting specifications capture all the safety requirements and improve the overall quality of the final specifications.

%The generate–evaluate loop continues until all safety requirements are covered and the specifications exhibit appropriate strength, or until a predefined iteration limit is reached, at which point the highest-scoring specification is forwarded to Kani for verification.

%As LLM-based assessments are inherently approximate, \speceval is designed as a lightweight semantic filter rather than a sound verifier, pruning low-quality specifications before formal verification. This evaluation-driven refinement helps ensure the resulting specifications comprehensively capture the target function’s safety concerns and improve the overall quality of the final specifications. 

% 虽然有llm，无法保证100%，但是能够缓解幻觉带来的影响，帮助提升spec生成的质量。实验xxx证实了这个观点。

%Due to the inherent uncertainty of LLM-based generation, there are many potential quality issues in the formal specifications produced by \specgen. First, generated specifications may fail to cover all extracted safety requirements. Second, the specifications may be vacuous, too strong, or too weak. 

% 这一段下一节再提
%Furthermore, during iterative specification generation and repair, when LLMs struggle to produce verifiable specifications, they may degenerate into generating trivially true constraints, circumventing verification at the cost of specification quality.

% TODO: an example that shows the refinement process in this step.

\subsection{Specification Verification}
\label{sec:specverify}

The \specverify agent invokes Kani to verify the target function under generated harnesses and specifications, and feeds the verification results back to \specgen for refinement when it fails. To improve the specification refinement process, \specverify constructs contextual feedback for \specgen based on the following three kinds of verification results from Kani:

\noindent{\textbf{Syntax errors}.}
If Kani reports syntax errors in the specifications, \specverify directly forwards the corresponding error messages to \specgen. These errors typically arise from incorrect usage of Kani primitives and can be resolved in the next iteration. 

\noindent{\textbf{Verification failures}.}
When the generated specifications are syntactically valid but fail to satisfy certain safety properties, the contextual feedback contains the following three parts.
First, it retrieves counterexamples from Kani for the failed properties. Second, when multiple harnesses are involved for a single target function, both passing and failing harnesses are retrieved for cross-reference. %support differential analysis. 
Third, if the failure falls into predefined failure types, the corresponding repair rule will be triggered. For example, when Kani reports violations related to the \code{assigns} clause, \specgen is explicitly prompted to refine the corresponding \code{\#[kani::modifies(...)]} annotations for mutable parameters. The full list of failure-specific repair rules can be found in our artefact.

\noindent{\textbf{Vacuous verifications}.}
A successful Kani verification does not exclude vacuous specifications
% \footnote{We use the term "vacuous" to denote specifications with over-constrained preconditions, under which the postconditions are never checked.}
, such as \code{\#[kani::requires(false)]}. 
Kani inherently checks the reachability of each Kani annotation, making vacuous specification detectable. However, it is tedious to check the reachability of all annotations from Kani's running log directly. In our implementation, \specverify interpolates a redundant \code{\#[kani::ensures(|\_| false)]} at the end of Kani postcondition. As long as the verification fails after the interpolation and succeeds before, the specification is considered non-vacuous.

% necessarily imply correct specifications, as LLMs may generate unsatisfiable preconditions, and they could cause postconditions to hold vacuously. 
% To detect such cases, \specverify appends an artificial postcondition \code{\#[kani::ensures(|\_| false)]} to the evaluated specifications before running Kani. If this postcondition is evaluated to be \code{false}, the preconditions are deemed unsatisfiable, and \specgen is instructed to regenerate the specifications. Otherwise, if only the artificial postcondition fails while all original properties hold, the verification is considered valid, indicating that the specifications meaningfully constrain program behaviour.

The refinement process, \ie, the \specgen-\speceval-\specverify loop, terminates when Kani verification succeeds, or the maximum iteration limit is reached. Then, the current specification is added to the candidate set. \tool allows users to set the target number of candidate specifications. 

% In our experiments, we set the candidate number to 2 and the iteration limit to 3.

%The interaction between \specverify and \specgen proceeds for a predefined number of iterations. If verification does not succeed within this limit, the loop terminates and defers to manual inspection to determine whether the failure is due to limitations of specification generation or genuine undefined behavior in the target function.

%To further improve the quality of generated specifications, we execute the generate–evaluate–verify loop multiple times, collect all successfully verified specifications, and select the best one. 

% \subsubsection{Evaluation with respect to user-intention.}
\subsection{Specification Selection}
\label{sec:specselect}
Given a set of specification candidates from \specverify, although all are valid and not vacuous, it is non-trivial to determine which candidate correctly formalises all desired safety properties, that is, fully satisfies users' intentions.
Intuitively, we tend to select the candidate with the weakest precondition and the strongest postcondition as the final result. However, it happens that the weakest precondition and the strongest postcondition are not from the same candidate.
For example, suppose we obtain two candidate specifications, $\{P\} ~C ~\{Q'\}$ and $\{P'\} ~C~ \{Q\}$, where $P' \implies P $ and $ Q \implies Q'$. The desired specification, however, is $\{P\} ~C~\{Q\}$. Although the two candidates are very close to the target, existing approaches~\cite{autospec,autoverus,llm_dafny,specgen} would typically start a new run of the tools, thereby missing the correct result that can be derived from the current candidate set. 

To address this challenge, we design a specification \emph{shuffle-and-implication strategy}, which shuffles predicates across candidates and selects the optimal combination of specification predicates.
If the best specification is already in the candidate set, the strategy will directly identify and return it in the first iteration. The remainder section introduces how our \emph{shuffle-and-implication strategy} effectively identifies the optimal specification.

\begin{algorithm}[t]
\caption{Specification Shuffle and Implication Strategy}
\label{alg:select-best-spec}

\KwIn{$F$: target function, $M$: max iterations,
$\mathcal{SP} = \{ (p_k, q_k, i_k) \mid k \in \mathbb{T} \}$: Specification set}
\KwOut{$(P_{best},Q_{best},I_{best})$: the optimal specification}

$i, j \gets 0; \mathbb{P} \gets getPreSet(\mathcal{SP})$\;
\While{$i < M$}{
    \textnormal{// get the weakest precondition} \\
    $p_w, i_w \gets p_k, i_k \;\text{s.t.}\; 
    \exists p_k \in \mathbb{P},\forall p_m \in \mathbb{P},~
    (p_m \Rightarrow p_k) \vee (p_k \nRightarrow p_m)$\;

    $\mathbb{Q} \gets getPostSet(\mathcal{SP})$\;

    \While{$j < M$}{
        \textnormal{// get the strongest postcondition} \\
        $q_s, i_s \gets q_k, i_k \;\text{s.t.}\;
        \exists q_k \in \mathbb{Q}, \forall q_m \in \mathbb{Q},~
        (q_k \Rightarrow q_m) \vee (q_m \nRightarrow q_k)$\;

        \If (\textnormal{  // no loop in $F$}) {$isEmptyInv(\mathcal{SP})$}{
            \If{$\textsc{KaniVerify}(F,p_w,q_s)==\textsc{Success}$}{
                \Return{$(p_w,q_s,None)$}\;
            }
        }
        \Else (\textnormal{  // has loop in $F$}){
            \If{$\textsc{KaniVerify}(F,p_w,q_s,i_w)==\textsc{Success}$}{
                \Return{$(p_w,q_s,i_w)$}\;
            }
            \ElseIf{$\textsc{KaniVerify}(F,p_w,q_s,i_s)==\textsc{Success}$}{
                \Return{$(p_w,q_s,i_s)$}\;
            }
            \ElseIf{$\textsc{KaniVerify}(F,p_w,q_s,i_w\wedge i_s)==\textsc{Success}$}{
                \Return{$(p_w,q_s,i_w\wedge i_s)$}\;
            }
        }

        $\mathbb{Q} \gets \mathbb{Q}\setminus\{q_s\}$\;
        $j$++\;
    }

    $\mathbb{P}\gets\mathbb{P}\setminus\{p_w\}$\;
    $i$++\;
}

\Return{$random(\mathcal{SP})$}\;
\end{algorithm}

\noindent{\textbf{Shuffle and implication strategy}}.
Algorithm~\autoref{alg:select-best-spec} presents details of the strategy. We take three inputs: the target function $F$, the maximum iteration times $M$, and a set of generated specification candidate tuples $\mathcal{SP}$ with the index set $\mathbb{T}$, where each element consists of a precondition, a postcondition, and a loop invariant. And then, we return the optimal specification tuple $(P_{best}, Q_{best}, I_{best})$. 

Within the maximum number of iterations, we select the weakest precondition and the strongest postcondition from the current pre- and post-condition sets (lines 4 and 8). If the function $F$ contains no loop, \ie all loop invariants in $\mathcal{SP}$ are none, we directly check the satisfiability of the current specification combination in Kani (line 10). If the verification succeeds, we return the current specification (line 11). Otherwise, we discard the current postcondition or precondition and test the next strongest or weakest combination in the subsequent iteration (lines 19 and 22). The two-layer nested loop guarantees that we will cover all possible combinations as long as $M \geq |\mathcal{SP}|$. In this case, there are at least $|\mathcal{SP}|$ satisfiable combinations, corresponding exactly to the original candidates in $\mathcal{SP}$. When no successful verification is found among the tested combinations, we randomly select a candidate from $\mathcal{SP}$ (line 23). This fallback occurs only when $M < |\mathcal{SP}|$.

% We first identify the weakest precondition $p_w$ and the strongest postcondition $q_s$ from $\mathcal{SP}$, and retrieve the remaining elements from their corresponding tuples (line 2-5). We then invoke Kani to verify the target function under ${p_w}$ and ${q_s}$. If the verification succeeds, $p_w$ and $q_s$ are selected as the final precondition and postcondition (line 6-7).

If the function $F$ contains loops, we test three possible loop invariants: $i_w$, $i_s$, and $i_w \wedge i_s$ for the selected $p_w$ and $q_s$ (lines 13-18), where $i_w$ is the invariant from the candidate tuple containing the current precondition $p_w$, and $i_s$ is from the tuple containing the current postcondition $q_s$. If none of these invariants make the verification succeed, we move on to the next iteration. Intuitively, the correct invariant is highly likely to be expressible in one of these three formats, as $i_w$ captures the assumptions required to enter the loop safely, while $i_s$ captures the conditions necessary to establish the desired postcondition after loop termination. Below, we provide a more formal explanation.

Consider two candidates $(P_1,Q_1, I_1)$ and $(P_2,Q_2,I_2)$, where $P_2 \implies P_1$ and $Q_2 \implies Q_1$, satisfiability of these candidates means that $P_1 \implies wp (F, Q_1)$ and $P_2 \implies wp (F, Q_2)$. If we focus on the loop in $F$, we obtain $P_1' \implies wp (while~(b)~invariant~I_1~\{S\}, Q_1')$ and $P_2' \implies wp (while~(b)~invariant~I_2~\{S\}, Q_2')$, where $P_1'$, $P_2'$, $Q_1'$, and $Q_2'$ are predicates immediately before and after the loop, calculated from $P_1$, $P_2$, $Q_1$, and $Q_2$, respectively.
By the \emph{while} and \emph{consequence} rule, we get $P_1' \implies I_1$, $I_1 \wedge \neg b \implies Q_1'$, and $I_1 \wedge b \implies wp(S,I_1)$, together with $P_2' \implies I_2$, $I_2 \wedge \neg b \implies Q_2'$, and $I_2 \wedge b \implies wp(S,I_2)$.

Our target is to find an invariant $I_{best}$, such that $P_1' \implies wp (while~(b)~invariant ~I_{best}~\{S\}, Q_2')$ holds.
Through the implication rules, we obtain the invariant based on the following rules:
%Based on the selected precondition and postcondition, we determine the best loop invariant $i_{best}$ from the generated candidates, without synthesizing new invariants. Our goal is to select $i_{best}$ such that the following rule is satisfied: 

% \begin{equation}\tag{Rule \theequation}
% \label{rule:best_inv}
% \frac{\{p_w\}\;\{i_{best} \land B\}\;S\;\{i_{best}\}\;\{q_s\}}
%      {\{p_w\}\;\text{while } B \text{ do } S\;\{q_s\}}
% \end{equation}

% We consider three cases, corresponding to lines 8–14 in Algorithm~\autoref{alg:select-best-spec}.
\begin{itemize}
    \item If $P_1' \Rightarrow I_2$ holds, as $I_2 \land \neg b \Rightarrow Q_2'$, select $I_{best} = I_2$.

    \item If $P_1' \Rightarrow I_1 \land I_2$ holds, as $I_2 \land \neg b \Rightarrow Q_2'$, select $I_{best} = I_1 \land I_2$.

    \item If $I_1 \land \neg b \Rightarrow Q_2'$ holds, as $P_1' \Rightarrow I_1$, select $i_{best} = I_1$.

    \item If none of the above cases apply, skip.
\end{itemize}

In the actual implementation, since we cannot easily derive $P_1'$, $P_2'$, $Q_1'$, and $Q_2'$, we instead directly check the satisfiability of the combined specification using each of the three candidate invariants. 

In our experiments, we set the candidate number to 2 and the maximum iterations to 2. 
% \autoref{sec:case_study} provides a case study, showing the effectiveness of the \emph{shuffle and implication strategy}.
% If $P_w \Rightarrow i_s$ holds, as $i_s \land \neg B \Rightarrow q_s$, ~\ref{rule:best_inv} holds. In this case, we select $i_{best} = i_s$.

% If $p_w \Rightarrow i_s \land i_w$ holds, as $i_s \land \neg B \Rightarrow q_s$, ~\ref{rule:best_inv} holds. In this case, we select $i_s \land i_w$ as $i_{best}$.

% If $i_w \land \neg B \Rightarrow q_s$ holds, as $p_w \Rightarrow i_w$, ~\ref{rule:best_inv} holds. In this case, we select $i_{best} = i_w$.

% If none of the above cases apply, we select any valid tuple as the final specification of the target function.

% We observe that this strategy can further improve the quality of the specification during the shuffle process. Intuitively,

% \subsubsection{Final Results}
% We cannot directly apply the automated specification evaluation scheme~\cite{DBLP:conf/fmcad/Lahiri24}, which evaluates specifications through mutated input-output pairs, since our specifications formalise safety properties of unsafe Rust code instead of functionality. XXXX more explanation...
% \subsection{Putting it all together}

\end{section}

\section{Evaluation}

In this section, we evaluate the effectiveness and efficiency of \tool in verifying unsafe Rust code with Kani. In summary, we seek to answer the following research questions:

\begin{itemize}
    % \item RQ1: How effective is \tool at generating specifications in verifying safety properties, compared with state-of-the-art approaches?
    \item RQ1: How effective is \tool at generating specifications in verifying safety properties? 
    \item RQ2: How does \tool compare with state-of-the-art specification generation approaches?
    \item RQ3: How effective is each component of \tool at generating valid specifications?
    % \item RQ3: How much time and money does \tool cost in verifying unsafe Rust functions across different steps of the \tool pipeline and different model settings?
    % \item RQ4: Can \tool identify potential vulnerabilities from failed verifications of unsafe Rust code?
\end{itemize}

\paragraph{Benchmarks.} 
We construct a benchmark of \ttfunc unsafe Rust functions, summarized in \autoref{tab:bench}. The benchmark consists of two parts: a gold set with ground-truth specifications (\dataa) and an unlabeled set without such specifications (\datab).

\begin{table}[t]
\caption{Summary of the benchmarks.}
\centering
\resizebox{0.65\linewidth}{!}{
\begin{tabular}{l | c | ccc | ccc}
\toprule
\multirow{2}{*}{Dataset} & \multirow{2}{*}{\#Task} &\multicolumn{3}{c|}{Function Lines} & \multicolumn{3}{c}{Documentation Lines} \\
\cmidrule(lr){3-5} \cmidrule(lr){6-8}
 & & min & max & avg & min & max & avg \\
\midrule
\dataa & 54 & 3 & 42 & 10.37 & 4 & 85 & 23.35 \\
% \midrule
% $D_{NGT}$ 
% \datab & \revise{70} & 3 & 55 & 15.41 & 1 & 55 & 15.77 \\
\datab & \revise{70} & 3 & 55 & \revise{15.60} & 4 & \revise{60} & \revise{15.69} \\
\bottomrule
\end{tabular}
}
\label{tab:bench}
\vspace{-1ex}
\end{table}

The \dataa consists of \stdfunc unsafe Rust functions from the verify-rust-std project~\cite{verify-rust-std} (commit: bacd51ca) and serves as the primary evaluation baseline. We select these functions for the following three reasons: 1) the safety of unsafe code in the Rust standard library is of central importance to the Rust community, making it a practically significant target for evaluation; 2) the safety requirements of the Rust standard library are comprehensive and of high quality defined by the \emph{rust-lang} developers; 3) and thanks to this project led by Amazon, where existing solutions provide human-expert-written specifications for these unsafe Rust functions, which perfectly satisfy the requirement of our benchmark as a reliable ground truth.

The \datab comprises \tpfunc unsafe Rust functions drawn from the Safe4U~\cite{safe4u} unsound samples, which collect 97 unsafe Rust functions along with their safety comments from the top 500 most downloaded libraries on \code{crates.io}. 
\revise{We exclude 27 functions due to Kani's (1) no support for \code{await} and inline assembly, (2) incompatibility with \code{packed\_simd}, and (3) inability to express contracts for functions that return mutable references to arguments.}

\paragraph{Implementation and Experimental Setup.}
\tool is implemented in a total of \ttpylines lines of Python code and \ttrslines lines of Rust code.  
\tool verifies programs in Kani 0.62.0. In experiments, we test \tool on three \revise{LLMs}: GPT-5 (gpt-5-2025-08-07), DeepSeek-v3.2 (2025-12-01), and Claude-Sonnet-4 (claude-sonnet-4-20250514). \revise{As these models have been widely used in prior work due to accessibility, moderate cost, and strong performance, ensuring reproducibility without specialized hardware.}
All experiments are conducted on a server running Ubuntu 24.04.1 LTS, with a 104-core Intel(R) Xeon(R) Gold 6230R CPU, 128 GB of RAM, and 2 TB of disk space.

\subsection{RQ1: Effectiveness of \tool}
\label{sec:rq1}

%% spec generation result: 
%%% std cases: 
% gpt-5: spec lines: max: 68, min: 1, avg: 21.24
% gpt-5: loop-inv lines: max: 2, min: 0, avg: 0.03
% deepseek-v3.2: spec lines: max: 21, min: 2, avg: 12.55
% deepseek-v3.2: loop-inv lines: max: 2, min: 0, avg: 0.1
% claude-4: spec lines: max: 48, min: 1, avg: 11.47
% claude-4: loop-inv lines: max: 8, min: 0, avg: 0.31
%%% non-std cases:
% gpt-5: spec lines: max: 57, min: 2, avg: 19.98
% gpt-5, loop-inv lines:  max: 15, min: 0, avg: 1.28
% deepseek-v3.2, spec lines: max: 69, min: 3, avg: 17.54
% deepseek-v3.2, loop-inv lines:max: 6, min: 0, avg: 0.42 
% claude-4, spec lines: max: 46, min: 2, avg: 13.46
% claude-4, loop-inv lines: max: 11, min: 0, avg: 0.93

\paragraph{Overall results.} We evaluate \tool on \dataa and \datab using three LLMs. Since Kani only checks whether a program satisfies the properties described in a specification, a specification that passes Kani may still be incomplete or even vacuous. Thus, we categorise the result into one of three types: \good, \bad, or \failure. A \good indicates that verification passes and the specification captures all safety requirements. A \bad indicates that verification passes, but the specification is incomplete or vacuous. A \failure means the verification process itself fails (e.g., due to a counterexample or syntax errors).
As there is no fully automated approach to distinguish \good from \bad cases, we manually performed the categorisation for \dataa, for which ground truth is available. 
\revisenew{Two co-authors, each with five years of experience in formal verification, independently evaluated the generated specifications in a blind review process, spending over 30 hours in total. We measured the agreement between their initial annotations using Cohen's $\kappa$~\cite{cohen_kappa}, with values ranging from 0.81 to 0.95 across different LLMs, as summarised in \autoref{tab:inter_rater_matrix}. The high $\kappa$ values and the few disagreements indicate strong inter-rater agreement. Disagreements were subsequently resolved through discussion until both co-authors reached a consensus.}
This manual evaluation is facilitated by Kani's concise specification primitives and clear semantics, which allow for a reliable comparison of semantic strength and equivalence.
Specifically, we manually assessed whether each generated pre- and post-condition is stronger than, weaker than, different from, or equivalent to the ground truth. A specification is labeled \good only if: (1) verification passes, (2) the generated precondition is equivalent to or weaker than the ground truth, and (3) the postcondition is equivalent to or stronger than the ground truth. If verification passes but these semantic criteria are not met, the result is labeled \bad. \autoref{fig:succ_bad_specs} shows examples of \good and \bad postconditions, respectively (analogous to the cases for preconditions), where the \tool-generated specification is shown on the left and the ground truth is on the right.

\begin{figure}[t]
    \centering
    \begin{subfigure}{\textwidth}
        \begin{subfigure}{0.49\textwidth}
            \input{_minted/tex/succ_spec/equv_ours}
        \end{subfigure}
        \hfill 
        \begin{subfigure}{0.49\textwidth}
            \input{_minted/tex/succ_spec/equv_gt}
        \end{subfigure}
        \caption{\revise{\emph{\Good} postcondition, equivalent to the ground truth.}}
    \end{subfigure}
    
    \vspace{1em}

    \begin{subfigure}{\textwidth}
        \begin{subfigure}[b]{0.58\textwidth}
            \input{_minted/tex/succ_spec/strong_ours}
        \end{subfigure}
        \hfill 
        \begin{subfigure}[b]{0.40\textwidth}
            \input{_minted/tex/succ_spec/strong_gt}
        \end{subfigure}
        \caption{\revise{\emph{\Good} postcondition, stronger than the ground truth.}}
    \end{subfigure}
    
    \vspace{1em}
    
    \begin{subfigure}{\textwidth}
        \begin{subfigure}{0.46\textwidth}
             \input{_minted/tex/bad_spec/weak_ours}
        \end{subfigure}
        \hfill 
        \begin{subfigure}{0.52\textwidth}
            \input{_minted/tex/bad_spec/weak_gt}
        \end{subfigure}
        \caption{\revise{\emph{Bad} postcondition, weaker than the ground truth.}}
    \end{subfigure}

    \vspace{1em}

    \begin{subfigure}{\textwidth}
        \begin{subfigure}[c]{0.54\textwidth}
            \input{_minted/tex/bad_spec/diff_ours}
        \end{subfigure}
        \hfill 
        \begin{subfigure}[c]{0.44\textwidth}
            \input{_minted/tex/bad_spec/diff_gt}
        \end{subfigure}
        \caption{\revise{\emph{Bad} postcondition, different from the ground truth.}}
    \end{subfigure}
    
    \caption{\revise{Examples of \good and \bad postconditions: \tool-generated (left) vs. ground truth (right).}}
    \label{fig:succ_bad_specs}
    % \vspace{-3ex}
\end{figure}

\begin{table}[ht]
\caption{\revisenew{
Summary of inter-rater agreement for the independent blind labeling of specifications generated by different LLMs that passed verification on \dataa. ``Both \textit{Good}'' and ``Both \textit{Bad}'' denote the number of specifications labeled as \textit{Good} and \textit{Bad} by both authors. ``Author A Only'' indicates the number of specifications labeled as \textit{Good} exclusively by Author A. ``Author B Only'' indicates the number of specifications labeled as \textit{Good} exclusively by Author B. }}

\centering
\resizebox{0.85\linewidth}{!}{
\begin{tabular}{c|c|c|c|c|c}
\toprule
\multirow{2}{*}{\revisenew{Model}} & \multicolumn{2}{c|}{\revisenew{Agreement}} & \multicolumn{2}{c|}{\revisenew{Disagreement}} & \multirow{2}{*}{\revisenew{Cohen's $\kappa$}} \\
\cmidrule(lr){2-3} \cmidrule(lr){4-5}
               & \revisenew{Both \textit{Good}} & \revisenew{Both \textit{Bad}} & \revisenew{Author A Only} & \revisenew{Author B Only} & \\
\midrule
\revisenew{GPT-5} & \revisenew{28} & \revisenew{16} & \revisenew{1} & \revisenew{3} & \revisenew{0.82} \\
\revisenew{DeepSeek-v3.2}  & \revisenew{27} & \revisenew{15} & \revisenew{0} & \revisenew{1} & \revisenew{0.95} \\
\revisenew{Claude-Sonnet-4} & \revisenew{32} & \revisenew{14} & \revisenew{2} & \revisenew{2}  & \revisenew{0.81} \\
\bottomrule
\end{tabular}
}
\label{tab:inter_rater_matrix}
\vspace{-2ex}
\end{table}

\autoref{tab:overall_res} presents the overall results. \tool performs comparable \good rates across three LLMs on \dataa, ranging from 51.85\% to 62.96\%, which demonstrates the effectiveness of the overall multi-agent design in \tool. Claude-Sonnet-4 and GPT-5 achieve slightly higher \good rates than DeepSeek-v3.2 on \dataa. The performance on \datab follows a similar trend. \revise{Note that, due to the lack of ground truth in \datab, passed cases on this dataset were not further classified into \good and \bad.} However, the \failure rate (\revise{28.57\%} to \revise{31.43\%}) is higher than that on \dataa (7.41\% to 20.37\%). A possible explanation is that function descriptions in \datab contain less information than those in \dataa, as shown in \autoref{tab:bench}. Since function descriptions directly influence the subsequent safety requirements and specification generations, lower-quality descriptions can lead to higher failure rates.

\begin{table*}[t]
\caption{Number of tasks verified by \tool across different LLMs. \revise{\emph{Pass} means that Kani verification passes, while} \emph{Failure} means that Kani verification fails to pass \revise{due to counterexamples or syntax errors}. \emph{\revise{Good}} means that Kani verification passes and the specification captures all \revise{safety requirements}. \emph{Bad} means that the verification passes but the specification \revise{is incomplete or vacuous}.}

\centering
% \small
\resizebox{0.85\linewidth}{!}{
\begin{tabular}{c|c|c|c|c|c}
\toprule
\multirow{2}{*}{Model} & \multicolumn{3}{c|}{\dataa Result} & \multicolumn{2}{c}{\datab Result} \\ \cline{2-6}
& \revise{\Good} & Bad & Failure & \revise{Pass} & Failure \\ \midrule
% GPT-5 & 31 (57.41\%) & 17 (31.48\%) & 6 (11.11\%) & 31 (70.45\%) & 13 (29.55\%) \\
% DeepSeek-v3.2 & 28 (51.85\%) & 15 (27.78\%) & 11 (20.37\%) & 30 (69.18\%) & 14 (31.82\%) \\ 
% Claude-Sonnet-4 & 34 (62.96\%) & 16 (27.78\%) & 4 (7.41\%) & 30 (69.18\%) & 14 (31.82\%)\\ \bottomrule
GPT-5 & 31 (57.41\%) & 17 (31.48\%) & 6 (11.11\%) & \revise{50 (71.43\%)} & \revise{20 (28.57\%)} \\
DeepSeek-v3.2 & 28 (51.85\%) & 15 (27.78\%) & 11 (20.37\%) & \revise{48 (68.57\%)} & \revise{22 (31.43\%)} \\ 
Claude-Sonnet-4 & 34 (62.96\%) & 16 (27.78\%) & 4 (7.41\%) & \revise{48 (68.57\%)} & \revise{22 (31.43\%)}\\ 
\bottomrule
\end{tabular}
}
\label{tab:overall_res}
% \vspace{-1ex}
\end{table*}
% gpt-5: pass: 31 + 19 = 50; fail: 13 + 7 = 20; pass rate: 50/70=71.43%, fail rate: 20/70 = 28.57%
% ds-v3.2: pass: 30 + 18 = 48, fail: 14 + 8 = 22; pass rate: 48/70=68.57% , fail rate: 22/70=31.43%
% claude-4: pass: 30 + 18 = 48, fail: 14 + 8 = 22; pass rate: 48/70=68.57% , fail rate: 22/70=31.43%

\autoref{tab:gen-results} summarises the pre-/post-condition and loop-invariant generation results across three LLMs on two datasets. GPT-5 consistently produces the largest specifications, averaging 18.44 lines \revise{per function} for \dataa and \revise{26.43} lines \revise{per function} for \datab, while DeepSeek-v3.2 and Claude-Sonnet-4 generate shorter specifications on average (9.56 and 11.96 lines for \dataa, respectively). Across both datasets, all models generate a few number of loop invariants. This arises for two reasons. First, LLMs may choose to generate only pre-/post-conditions even for functions containing loops, and Kani can still verify these functions via unwinding. Second, Kani currently supports loop-invariant primitives only for \texttt{while} loops. Other loop forms must be manually rewritten in a consistent manner, which is error-prone. In our experiments, we let Kani perform full unwinding for non-\texttt{while} loops by default, and impose a bounded unwind only when full unwinding would be too costly.

\begin{table}[t]
\caption{Summary of generated specifications in \tool across different LLMs.}
\centering
\resizebox{\linewidth}{!}{
\begin{tabular}{c|ccc|ccc |ccc |ccc}
\toprule
\multirow{1}{*}{Dataset} & \multicolumn{6}{c}{\dataa} & \multicolumn{6}{c}{\datab} \\
\cmidrule(lr){1-7} \cmidrule(lr){8-13}
\multirow{2}{*}{Lines} & \multicolumn{3}{c|}{pre- and post-conditions} & \multicolumn{3}{c|}{loop invariants} & \multicolumn{3}{c|}{pre- and post-conditions} & \multicolumn{3}{c}{loop invariants} \\
\cmidrule(lr){2-4} \cmidrule(lr){5-7} \cmidrule(lr){8-10} \cmidrule(lr){11-13}
& min & max & avg & min & max & avg & min & max & avg & min & max & avg \\
\midrule
GPT-5 & 1 & 68 & 18.44 & 0 & 2 & 0.04 & \revise{1} & \revise{97} & \revise{26.43} & 0 & 15 & \revise{1.20} \\
DeepSeek-v3.2 & 2 & 21 & 9.56 & 0 & 2 & 0.07 & \revise{1} & 69 & \revise{13.54} & 0 & \revise{7} & \revise{0.73} \\
Claude-Sonnet-4 & 1 & 48 & 11.96 & 0 & 8 & 0.15 & \revise{1} & \revise{60} & \revise{15.56} & 0 & \revise{12} & \revise{1.04} \\
\bottomrule
\end{tabular}
}
\label{tab:gen-results}
% \vspace{-2ex}
\end{table}

\begin{center}
  \begin{minipage}[t]{0.56\textwidth} 
    \vspace{0pt} 
    \paragraph{Invocation Times.} \autoref{fig:invocation-box} shows the invocation times of each component in \tool on \dataa and \datab. As mentioned in \autoref{sec:approach}, we set the maximum invocation times of \safetyreq to 3, \specgen-\speceval loop to 5, and \specverify to 3. Overall, Claude-Sonnet-4 has the lowest median invocation times and the smallest variance across all stages. For \specgen and \speceval, although the highest invocation times reach the predefined threshold in GPT-5, more than 50\% of tasks terminate in 5 iterations across all three models.
  \end{minipage}
  \hfill
  \begin{minipage}[t]{0.43\textwidth} 
    \vspace{0pt}
    \centering
    \includegraphics[width=0.95\textwidth]{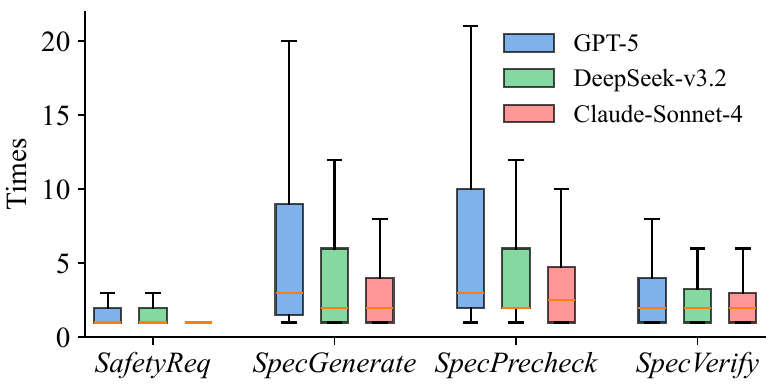}
    \vspace{-1.5pt}
    \captionof{figure}{Invocation times of \tool components.}
    \label{fig:invocation-box}
  \end{minipage}
\end{center}

% \noindent{\textbf{Money Costs}}. 
\paragraph{Money Costs.}\autoref{tab:token-cost} shows the money \revise{and token} cost on \dataa and \datab using different LLM models. 
At the time of our experiments, the cost per 1 million input (output) tokens is \$1.25 (\$10) for GPT-5, \$0.29 (\$1.14) for DeepSeek-v3.2, and \$3 (\$15) for the Claude-4-Sonnet. DeepSeek-v3.2 incurred the lowest overall cost due to its very low per-token price. Claude-4-Sonnet, despite having a higher per-token cost than GPT-5, had a lower overall cost because its invocation times were fewer. \revise{Apart from that, Claude-Sonnet-4 maintains the most concise token profile. This high efficiency in token usage effectively compensates for its premium unit pricing, resulting in higher cost-effectiveness compared to GPT-5.}

\begin{table}[htbp]
   \caption{Comparison of average money and token cost per function.}
    \centering
    \resizebox{0.65\linewidth}{!}{
    \begin{tabular}{c|c|c|c}
        \midrule
        LLMs & GPT-5 & DeepSeek-v3.2 & Claude-Sonnet-4 \\
        \hline
        Avg. Cost (USD) & 0.41 & 0.02 & 0.26 \\
        Avg. Tokens & 219,218 & 90,116 & 82,424 \\
        \bottomrule
    \end{tabular}
    }
    \label{tab:token-cost}
\vspace{-10pt}
\end{table}

\subsection{RQ2: Comparisons}
\label{sec:rq_comparison}
Since there is no prior work on using LLMs to generate specifications for unsafe Rust verification in Kani, we select AutoSpec~\cite{autospec} as the closest point of comparison. We delay the detailed discussion of the reason for choosing AutoSpec for comparison to \autoref{sec:discussion}.

To enable a fair comparison, we adapt AutoSpec from Frama-C to Kani through three steps. First, we convert all Frama-C-specific prompts into Kani-compatible prompts. For example, we replace "assuming you are a Frama-C expert" with "assuming you are a Kani expert", replace Frama-C-based examples with Kani examples that we have in \tool, \etc. Second, we replace Frama-C API calls with Kani commands that invoke the verifier to check generated specifications and collect running results. Third, we adapt the whole error message processing component to make it compatible with Kani's outputs. Throughout this adaptation, we preserve AutoSpec's overall pipeline, including its use of LLMs to generate candidate specifications, its iterative construction of specifications, and its specification repair approach based on error-message types.

\autoref{tab:cmp-autospec} shows the performance of \tool against AutoSpec in terms of specification quality on \dataa benchmark.
Overall, \tool produced approximately 82\% more \good specifications than AutoSpec, reflecting the effectiveness of its specification evaluation and logical implication–based selections in eliminating low-quality candidates.
In particular, \tool substantially reduced unsuitable preconditions, which include both incorrect and overly restrictive ones, by about 74\% compared with AutoSpec. A similar trend was observed for postconditions. \tool generated around 22\% fewer incorrect or overly weak postconditions, indicating better semantic alignment with the intended safety requirements.

Moreover, during specification verification, \tool provides LLMs with counterexamples as well as both failing and successful harnesses, together with error-specific fixing strategies, enabling issues to be identified and repaired early in the pipeline. As a result, \tool reduced verification failures by approximately 57\% relative to AutoSpec. These results demonstrate that \tool not only increases the overall success rate of specification generation, but also significantly improves the semantic quality and reliability of the generated specifications.
% 因为\tool有spec precheck对spec quality进行打磨，加上我们最终使用logical implication对spec进行筛选，我们最终生成的success spec 远多于autospec（31 vs. 17）。具体而言，不合适的 pre （#wrong pre + #stronger pre), \tool生成了7个，而autospec生成了27个。不合适的post (#wrong post + #weaker post)，\tool生成了14个，autospec生成了18个。
% 此外，\tool 在spec verify中，包含针对不同错误的处理策略，使得syntax errors及时得到修正，最终failed cases也远少于autospec (6 vs 14)。

\begin{table*}[t]
    \caption{Performance of \tool against AutoSpec (direct prompting) under GPT-5.}
    \centering
    \resizebox{\textwidth}{!}{
    \begin{tabular}{c|c|c|c|c|c|c|c|c|c|c|c|c}
        \toprule
         \multirow{2}{*}{Tool} & \multirow{2}{*}{\#Tasks} & \multicolumn{3}{c|}{Results} &\multicolumn{4}{c|}{Precondition} &  \multicolumn{4}{c}{Postcondition} \\ 
         %\cline{2-11}
          & & \Good & Bad & Failure & Weaker & Equiv & Wrong  & Stronger  & Stronger	& Equiv	& Weaker &	Wrong \\ \midrule
    \tool    &       54      &   31   &    17     &    6     &    3    &   38   &   0      &    7     &    16 & 18 & 13 & 1    \\
    AutoSpec   &       54      &    17  &    23     &    14     &    3    &   24   &     13    &    14     &    28 & 8 & 11 & 7    \\ \midrule
   % Percentage & - & \uparrow 82.35\%  & \downarrow 26.09\% & \downarrow 57.14 & \multicolumn{2}{c|}{\uparrow 51.85\%} & \multicolumn{2}{c|}{\downarrow 74.07\%} & \multicolumn{2}{c|}{\downarrow 5.56\%} & \multicolumn{2}{c}{\downarrow 22.22\%} \\ 
 Surpass & - & $\uparrow$ 82\%  & $\downarrow$ 26\% & $\downarrow$ 57\% & \multicolumn{2}{c|}{$\uparrow$ 52\%} & \multicolumn{2}{c|}{$\downarrow$ 74\%} & \multicolumn{2}{c|}{$\downarrow$ 6\%} & \multicolumn{2}{c}{$\downarrow$ 22\%}  \\
    \bottomrule
    \end{tabular}
    }
    \label{tab:cmp-autospec}
\vspace{-10pt}
\end{table*}

\subsection{RQ3: Ablation Study}
\label{sec:rq_ablation}
To understand the effectiveness of each essential component in \tool, we conduct three ablation studies, and the results are shown in \autoref{tab:ablation}. In this study, we compare the full \tool with variants that respectively exclude the \safetyreq, \speceval, \specverify agents, \revise{as well as the few-shot examples in \specgen. Furthermore, we compare \tool against a naive LLM prompting baseline (i.e., removing all agents)}.
We record the \good rate, \pass rate, \bad rate, and \failure rate over the same set of 54 verification tasks that we used in \autoref{sec:rq1} for the comparison with AutoSpec. For simplicity, we perform the ablation study under GPT-5.

% The results in \autoref{tab:ablation} show that the full \tool setting achieves the best overall performance, successfully verifying 88.89\% (48 out of 54) while maintaining the lowest rate of Bad and Fail cases. We observe that ...

\noindent{\textbf{\tool w/o \safetyreq.}} We replaced the \safetyreq agent with a simple LLM that generates the safety requirement list directly from the function's documentation, which is then passed to subsequent agents. Without quality checks and feedback-driven refinement, the resulting safety requirement lists were often misaligned with the original safety concerns of the function, making it difficult to produce specifications with appropriate strength. Experimental results show that this simplification reduced the \revise{\good rate of specification generation} by approximately 46.30\%, and the \bad rate increased by around 3.70\%.

\noindent{\textbf{\tool w/o \speceval.}} We removed the \speceval agent from \tool, delivering the specifications generated by \specgen directly to \specverify. Without the quality control provided by \speceval, the generated specifications are more likely to be inaccurate and misaligned with the intended safety requirements. The experimental results confirm that removing this agent increased the proportion of specifications with inappropriate strength, such as overly strong preconditions or overly weak postconditions, by 12.96\%, and the \good rate dropped by 35.19\%.

\noindent{\textbf{\tool w/o \specverify.}} We replace the \specverify agent with a simplified agent that forwards pure Kani error messages to \specgen for the refinement, without contextual feedback mentioned in \autoref{sec:specverify}, including counterexamples, failing and successful harnesses, error-specific fixing strategies, and the specification selection strategy in \autoref{sec:specselect}. The experimental results show that removing these components substantially degrades performance. The \failure rate increased by 29.63\%, and the \good rate decreased by 35.19\%. Without this detailed feedback, errors in Kani specifications cannot be corrected effectively, leading to more iterations in the generate–evaluate–verify loop and, ultimately, exceeding the maximum iteration limit, which causes generation to fail.
% 37.04\% -> 35.19\%

\noindent{\revise{\textbf{\tool w/o few-shots.}}}
\revise{We removed the few-shot examples from \specgen, meaning that the agent could no longer retrieve similar function-specification pairs to serve as in-context references. Without these few-shot examples, the agent lacks concrete demonstrations of how to map specific safety requirements and code patterns to Kani's primitives, making it difficult to maintain the required precision and syntax for various safety properties. Removing few-shot examples increases \failure rate by 12.96\% and reduces \good rate by 37.04\%, showing that few-shot examples are critical guidance for encoding Kani-specific safety constraints, enabling more accurate specifications.}

\noindent{\revise{\textbf{Naive LLM prompting.}}}
\revise{We also implement a baseline prompting strategy for comparison. Specifically, we replace \tool's multi-agent architecture with a direct prompting approach, using Kani as an oracle to provide error feedback for iterative refinement (up to 10 rounds). Experimental results show that this simplified approach significantly degrades performance. The \failure rate increases by 37.04\%, while the \good rate drops by 42.59\%. These results confirm that \tool's performance gains stem from its specialized multi-agent design rather than iterative prompting alone.}

\begin{table}[t]
    \caption{Ablation study in GPT-5.}
    \centering
    \resizebox{0.9\textwidth}{!}{
    \begin{tabular}{c|l|c|c|c|c}
        \toprule
         NO.  & Setting & \Good & Bad & Pass & Failure \\ \midrule
         1 & Full \tool & 31 \xspace 57.41\% & 17 \xspace 31.48\% & 48 \xspace 88.89\% & 6 \xspace 11.11\% \\
         2 & \tool w/o \safetyreq &  6 \xspace $\downarrow$ 46.30\% & 19 \xspace $\uparrow$ 3.70\%  &  25 \xspace $\downarrow$ 42.59\% &29 \xspace $\uparrow$ 42.59\% \\ 
         3 & \tool w/o \speceval &  12 \xspace $\downarrow$ 35.19\% & 24 \xspace $\uparrow$ 12.96\% & 36 \xspace $\downarrow$ 22.22\% & 18 \xspace $\uparrow$ 22.22\% \\ 
         4 & \tool w/o \specverify & 12 \xspace $\downarrow$ 35.19\% & 20 \xspace $\uparrow$ 5.56\% & 32 \xspace $\downarrow$ 29.63\% & 22 \xspace $\uparrow$ 29.63\% \\ 
         5 & \tool w/o few-shots & 11 \xspace $\downarrow$ 37.04\% & 30 \xspace $\uparrow$ 24.07\% & 41 \xspace $\downarrow$ 12.96\% & 13 \xspace $\uparrow$ 12.96\% \\
         6 & Naive LLM prompt & 8 \xspace $\downarrow$ 42.59\% & 20 \xspace $\uparrow$ 5.56\% & 28 \xspace $\downarrow$ 37.04\% & 26 \xspace $\uparrow$ 37.04\% \\ 
         \bottomrule
         
    \end{tabular}
    }
    \label{tab:ablation}
    \vspace{-2ex}
\end{table}

\subsection{Case Study}
\label{sec:case_study}
\noindent{\textbf{Case Study 1}}. 
This case study compares \tool and AutoSpec~\cite{autospec} on generating specifications for the same function, \code{offset}, whose signature and documentation are shown in \autoref{fig:desc-req}. AutoSpec generates specifications directly from the function implementation and produces preconditions as shown in \autoref{fig:autospec-tool}. Although syntactically correct, these preconditions are semantically redundant, all expressing the same constraint that "the computed offset \code{count * size\_of::<T>()} must not overflow \code{isize}". But Kani verification fails with the error "Offset result and original pointer must point to the same allocation", revealing a missing safety property that "the entire memory range between the original pointer and the resulting pointer must lie within the bounds of the same allocated object" in the specification. Since this constraint is not explicitly stated in the description, AutoSpec cannot capture this safety requirement for the pointer computation.

% todo: highlight req3
\begin{figure}[t]
    \begin{subfigure}[t]{0.51\textwidth}
        \input{_minted/tex/desc.tex}
    \end{subfigure}
    \hfill
     \begin{subfigure}[t]{0.46\textwidth}
        \input{_minted/tex/safety-req.tex}
    \end{subfigure}
    \caption{\code{offset}'s description (left) and \safetyreq-generated safety requirement list (right).}
    \label{fig:desc-req}
    \Description{}
\end{figure}

In contrast, \tool derives specifications from a structured safety requirement list extracted from the function documentation, as shown in \autoref{fig:desc-req}. The \safetyreq agent iteratively refines this list to ensure that all relevant safety concerns are covered while avoiding redundant or overlapping requirements. Guided by this refined list, \tool generates a Kani specification that explicitly encodes the missing constraint using \code{kani::mem::same\_allocation} as highlighted in \autoref{fig:autospec-tool}, accurately capturing the requirement that both pointers belong to the same allocation. As a result, the generated specification successfully passes Kani verification.

This case study highlights a key limitation of code-centric specification generation. Safety constraints that are implicit in documentation but not directly reflected in code are easily overlooked. By contrast, grounding specification generation in a comprehensive and non-redundant set of documentation-derived safety requirements enables \tool to produce accurate specifications.

\begin{figure}[t]
    \begin{subfigure}{0.50\textwidth}
            \input{_minted/code/autospec_spec.m}
    \end{subfigure}
    \hfill
     \begin{subfigure}{0.49\textwidth}
            \input{_minted/code/kapilot_spec.m}
    \end{subfigure}
    \caption{Preconditions generated by AutoSpec (left) and \tool (right).}
    \label{fig:autospec-tool}
    \Description{}
    \vspace{-1ex}
\end{figure}

\noindent{\textbf{Case Study 2}}.
This case study demonstrates how \tool identifies and repairs subtle specification defects in \specverify. The Rust function \code{as\_flattened}, as shown in~\autoref{fig:false-check}, is parameterised by a generic type \code{T}, which may be either zero-sized (ZST) or non-zero-sized. \autoref{fig:desc-req} shows the safety requirement list extracted by the \safetyreq agent. In the first specification generation attempt, the LLM-produced precondition is shown in \autoref{fig:false-check} line 2. This precondition implicitly assumed \code{T} to be zero-sized and failed to account for the non-ZST case. Consequently, when \code{T} is non-zero-sized, the precondition evaluates to \code{false}, causing postconditions to be vacuous, allowing verification to succeed without imposing any meaningful behavioural constraints.

To automatically detect such vacuous specifications, \tool applies a vacuity check by appending \code{\#[kani::ensures(|\_| false)]} to the postcondition. If the precondition collapses to \code{false}, this new postcondition is incorrectly verified as \code{true}, signaling the presence of vacuity. Using this mechanism, \tool identifies the flawed precondition and feeds explicit feedback to \specgen agent. The LLM is then instructed to regenerate the precondition while accounting for both ZST and non-ZST cases. The corrected precondition from the second iteration is shown in \autoref{fig:false-check} line 3.

This case highlights that syntactically correct and verified specifications from LLMs may still be semantically meaningless due to vacuous assumptions. It is difficult to detect via manual inspection. By systematically enforcing vacuity checks, \tool ensures that generated specifications are non-trivial, significantly improving the reliability of automated specification generation.

% This case highlights a critical failure mode of LLM-generated specifications. Syntactically correct and verifiable specifications may still be semantically meaningless due to vacuous assumptions. Such issues are difficult to detect through manual inspection. By systematically enforcing vacuity checks, \tool ensures that generated specifications are non-trivial, significantly improving the reliability of automated specification generation.

\begin{figure}[t]
    \centering
    \input{_minted/code/false_check.m}
    \caption{Specifications generated for \code{as\_flattened} based on the safety requirements. The prefixes "-" and "+" denote, respectively, an incorrect precondition generated first and a correct one generated after correction.}
    \label{fig:false-check}
    \vspace{-1ex}
\end{figure}

% \noindent{\textbf{Case Study 3}}. TODO: illustrate the effectiveness of the spec selection algorithm.
\section{Discussion}
\label{sec:discussion}

% \subsection{More Large Language Models}
% 为啥只测了这几个model，没有测更多？为什么只是这几个版本？
% 木有那么多钱，现有实验已经证明了方法有效性，论文目的不是对比大模型性能，而是我们的工具好
% 我们的实现可以很容易用其他的任何模型，transportability

\noindent{\textbf{Why comparing with AutoSpec.}}
AutoSpec is a similar LLM-based specification generation framework designed for Frama-C~\cite{framac}, an automated verification tool whose interface is similar to Kani. In both frameworks, specifications are expressed as annotations embedded in the source code. Moreover, AutoSpec has a classical and representative pipeline for specification generation tasks that recursively invokes LLMs till the verification succeeds. Although prior works~\cite{autoverus, alphaverus} have utilised LLM in Rust verification on Verus, the workflow for Verus automation cannot be directly adapted to Kani. Thus, we choose AutoSpec for comparison.
% RQ里已经提过了，但是可能有疑问，反正都改了代码，autoverus一系列工作也可以比较，其他的太早了，llm dafny
% 1. Autospec作用相当于baseline，代表一系列的LLM spec gen的工作
% 2. verus 工作流程和kani区别太大，改autoverus不适用
% 3. llm dafny也是ATP，都很粗糙

\noindent{\textbf{Evaluating specification quality.}}
An important question is whether a specification that successfully passes Kani verification truly and precisely captures the intended safety properties. In our evaluation, assessing this aspect still partially relies on manual inspection, and we do not yet provide a fully automated solution. However, \tool incorporates three design choices to improve specification quality. First, \speceval uses LLMs to validate alignment between generated specifications and documented safety requirements. Second, \specverify explicitly filters out unreachable results, preventing vacuous proofs caused by overly restrictive preconditions. Third, shuffle-and-implication re-evaluates pre- and post-conditions across candidates to explore a broader constraint space, increasing the likelihood that the resulting specification captures a wider range of critical memory-safety properties.

Prior work~\cite{mut-in-out-pair} evaluates the correctness and completeness for post-conditions by mutating implementation code or input–output pairs~\cite{mut-in-out-pair}. These mutations are intended to alter functional results, and if the post-conditions still hold, they are considered too weak. However, our focus is on panic and memory safety properties rather than pure functional correctness. Changes in functional results do not necessarily indicate memory safety violations or runtime panics. The specification ranking~\cite{he2025ranking} approach designs four rating criteria, but they are for P models~\cite{pmodel}. Therefore, existing approaches are not well-suited to our setting. 

\noindent{\revise{\textbf{Effects of documentation quality.}}}
\revise{Poor documentation may lead to missing or underspecified constraints, potentially resulting in bad specifications. Failures stem from (i) reliance on incomplete examples omitting corner cases, exceptional paths, or implicit assumptions, and (ii) ambiguities and inconsistencies across cross-referenced documentation. These are orthogonal to \tool and can be mitigated by preprocessing steps such as LLM-based ambiguity resolution~\cite{req_ambiguity_detect,llm_for_req_eng} or extracting precise structured constraints~\cite{auto_software_doc}. We leave this as future work.}

% 即使通过了kani证明，是否精准描述了想要的properties？
% 实验依旧依赖人工检查，没有提出自动的检查方案，但是
% kapilot流程中3处设计为这个问题：
% spec precheck利用LLM检查spec是否符合safety requirements; 
% spec verify的false, unreachable检查，防止了vacuous情况
% shuffle pre- post-condition最大程度上增强了spec的覆盖范围，尽可能涵盖最多的properties
% 

% 第二段：input-output exmaples方法不行。比如，在panic例子，XXX

% \subsection{Why Kani?}
% 安全properties通常抽象，不易从代码中直接提取，没有标准定义，也没有标准形式化表述

% Kani 里面定义了一些安全UB的postcondition，目前支持最好
% 

% How different are functional correctness verification and safety verification?

% future work 1
% improve kani, make CBMC support unbounded checking?

%% Undefined Behavior
% 依赖于Kani, 所以能够检测到的UB受限于Kani。

%% limitations
\noindent{\textbf{Expressiveness of specifications}.
\tool relies on Kani to express specifications. Although Kani provides specification primitives for memory safety and runtime correctness, its expressiveness is limited for certain natural-language safety requirements. For instance, lifetime-related properties (\eg, "For the entire lifetime \code{'a} of the returned \code{\&'a CStr}, the memory region starting at \code{ptr} and spanning the C string must not be mutated.") lack direct support, forcing approximation of the intended semantics in specifications. 
In addition, semantic overlap exists among Kani primitives. For instance, both \code{mem::checked\_align\_of\_raw} and \code{mem::can\_dereference} capture alignment properties, which may lead LLMs to generate redundant constraints and introduce unnecessary specification verbosity. The performance of \tool is bounded by the expressiveness of Kani primitives.

\noindent{\textbf{Bounded verification in Kani}}.
Kani is based on bounded model checking and is limited to provide fully unbounded proof. While recent support for loop invariants improves scalability, it is currently restricted to \code{while} loops, requiring manual rewriting for other loop forms. Moreover, nondeterministic modelling of vector-like data structures still relies on concrete capacity bounds, limiting the expressiveness of unbounded abstractions. Addressing these limitations is an important direction for future work.

% kani的表述能力
% kani primitives表述能力不足：有些nl safety req，没有适当的kani primitives来表达，比如：`For the entire lifetime 'a of the returned `&'a CStr`, the memory region starting at `ptr` and spanning the C string must not be mutated.`，来自c_str.rs, 305，fn from_ptr. 没有直接的表述lifetime的primitives. 导致spec eval评估分数低，spec eval <-> spec gen 之间的loop迭代次数太多。
% kani primitives功能不单一，可能与其他primitives存在重复的表述能力。可导致spec书写冗余。e.g., kani::can_dereference(ptr)，包含了 ptr is aligned的性质，kani::mem::checked_align_of_raw(ptr)也是check alignment的性质。llms可能生成两者的 &&，导致冗余。实际上，只需要一个kani::can_dereference(ptr)就可以了。例子：non_null.rs, 1263, drop_in_place

% kani bounded proof
% kani是一个bouned model checking tool，recent versions提供了loop invariants的prmitives，不需要fully unwind loops，提升了verification scalabilty。但是，仍然不能给提供fully unbouned proof。
% 目前kani v0.62.0，仅支持针对while-loop写loop invariants。对于surface rust中的其他形式的loops，可以手动改写成while loop。此外，对于一些vector-like的类型，使用kani 进行nondet时，仍然需要指定一个concrete value as limited capacity。比如vec![0, N]，仍然需要一个bounded N。后续可以修改kani，支持多维度的nondet。

\noindent{\textbf{Threat to validity.}}
The \datab dataset poses no risk of data leakage, as it contains no existing specifications. For the functions in \dataa with human-written ground truth, a potential risk exists because the models’ training data are undisclosed.
However, the knowledge cutoffs of GPT-5, Claude-Sonnet-4 and DeepSeek-v3.2 are September 2024, March and September 2025, respectively, at which time Kani specifications were scarce. Moreover, during our evaluation, we did not observe any non-trivial cases where the generated specifications are the same as the ground truth. We therefore believe the threat is minimal.

% \paragraph{Data Leakage}
% % gpt-5 24/09, kani spec数据集非常稀少，没有威胁，完全一样的比例很少

\section{Related Work}

\subsection{Verification of unsafe Rust code}
% Foundation of safety verification and related tools

% Gillian-Rust, VeriFast

% Verus (semi-automatic, functional correctness, XXX Rust)

% Need to answer: why choose Kani

Several studies have focused on unsafe Rust verification.
RustBelt~\cite{rustbelt} establishes a semantic typing framework in Coq~\cite{coq} based on lifetime logic to prove the type soundness of a core subset of Rust and ensure that safe APIs correctly encapsulate unsafe code. RustHornBelt~\cite{rusthornbelt} extends this foundation with parametric prophecies to enable functional correctness verification using first-order logic specifications. RefinedRust~\cite{refinedrust} builds on refined ownership types to support automated reasoning for both safe and unsafe Rust and produces machine-checkable proofs. GillianRust~\cite{gillianrust} combines lifetime logic with SMT-based symbolic execution to reason about unsafe code. These approaches typically require substantial manual effort, including modeling and auxiliary proof construction. Verus~\cite{verus} adopts an SMT-based verification approach over Rust-like programs, supporting safe Rust and limited unsafe constructs, but still relies on user-provided proof annotations. Automated approaches based on bounded model checking~\cite{kani, smack, seahorn} and symbolic execution~\cite{rvt, crux, panic-checker-klee} have also been applied to unsafe Rust. Kani~\cite{kani} uses bit-precise bounded model checking to verify both safe and unsafe Rust, supporting function contracts and loop invariants for modular verification. UnsafeCop~\cite{unsafecop} extends Kani with loop bound inference, loop stubbing, and scheduling strategies to improve scalability. SMACK~\cite{smack} translates LLVM bitcode into Boogie IR~\cite{boogie} to detect memory safety violations. Seahorn~\cite{seahorn} performs model checking and abstract interpretation at the LLVM IR level, while RVT~\cite{rvt} extends KLEE to identify memory safety bugs via symbolic execution.

% 自动
% good at memory safety 
% 

% Verus, Prusti for safe Rust only
% unsafe rust: interactive prover
% Kani, automatic, 
% Kani in large-scale verification

\subsection{LLM for formal verification}
% AutoSpec, SpecGen, LLM Dafny, 
% \paragraph{LLM-based specification generation.}
Several studies~\cite{autospec, specgen, pei-llm-reason-loop-inv, rank-llm-gen-loop-inv} focus on automatically generating formal specifications using LLMs.
AutoSpec~\cite{autospec} takes the source code as input, employing a call-graph-based hierarchical decomposition and strategically inserts placeholders into the code to guide LLMs in generating candidate specifications through iterative, bottom-up inference. These candidates are then formally validated by a prover, with invalid ones being filtered out. The framework repeats this generate-validate cycle, progressively refining the specification set until verification succeeds or an iteration limit is reached. SpecGen~\cite{specgen} is another LLM-driven approach for specification generation. It first engages the LLM to produce candidate specifications. When the initial candidates fail verification, it applies mutations and employs a weighted heuristic to select promising variants for validation. Similar to AutoSpec, it derives specifications directly from the source code.
Beyond direct generation, several studies address challenges in improving the quality and selection of LLM-generated invariants. Pei et al.~\cite{pei-llm-reason-loop-inv} fine-tune LLMs using a scratchpad-style prompting strategy to enable stepwise reasoning for invariant inference, achieving performance comparable to dynamic analysis tools without relying on execution traces. Chakraborty et al.~\cite{rank-llm-gen-loop-inv} propose a learning-based ranking approach that prioritises inductive loop invariants based on semantic similarity to the verification task, reducing the number of expensive verifier invocations.

% \paragraph{LLM-assisted proof and verified code generation.}
Another category of work integrates LLMs with verification tools to generate proof annotations and verified code.
%AutoVerus
AutoVerus~\cite{autoverus} combines domain expertise with formal methods to assist LLM agents in generating, refining and debugging proof annotations for Verus programs. By iteratively leveraging verification feedback, it achieves a high success rate in producing correct proofs.
% SAFE
SAFE~\cite{safe} addresses data scarcity in verified code generation through a self-evolving framework that combines synthetic data generation with model fine-tuning, demonstrating improved efficiency and accuracy compared to approaches that rely solely on general-purpose LLMs. 
% AlphaVerus
AlphaVerus~\cite{alphaverus} extends this idea by introducing a fully automated self-improvement loop that integrates candidate exploration, tree-search-based repair, and specification critique, enabling the system to bootstrap its verification capabilities without human intervention. 
Similarly, the work~\cite{llm_dafny} synthesises Dafny~\cite{dafny} implementations together with formal specifications from concise functional descriptions using multiple prompting strategies, and evaluates the results using the Dafny verifier. These approaches primarily target functional correctness, rather than the low-level memory safety constraints required for verifying unsafe Rust code.
\section{Conclusions}
We presented \tool, a fully automated multi-agent LLM framework for generating Kani specifications to verify unsafe Rust code. \tool distills documented safety concerns into structured safety requirements that guide specification generation, and refines candidate specifications through a generate–precheck–verify loop, followed by a shuffle and implication strategy to determine the final specifications.
We evaluated \tool on \ttfunc Rust functions. It successfully generated specifications for \NGtPassRate of functions without ground truth and for \GtPassRate of functions with ground-truth specifications, with \GtSuccRate of the generated specifications being semantically equivalent to or better than the ground truth. Compared with AutoSpec, \tool improves the success rate of generating verifiable specifications by \PassRateMore and equivalent-or-better specifications by \SuccRateMore.

\section{Data Availability}
% The artefact includes the source code and prompts of \tool and the scripts to reproduce the experiments. The artefact is uploaded as supplementary material for reviewing. We will make it public on GitHub upon acceptance.
% We make the source code of \tool, dataset, and scripts to run experiments available at a public repository {xxx}.
All data and scripts are publicly available at \cite{kapilot_artifact}.

\bibliographystyle{ACM-Reference-Format}
\bibliography{ref}

@online{ublist,
  author =       "The Rust Reference",
  year =         "2026",
  title =        "The Rust Reference",
  url =          "https://doc.rust-lang.org/reference/behavior-considered-undefined.html",
  lastaccessed = "Janurary 20, 2026",
}

@online{rustbook,
  author =       "The Rust Book",
  year =         "2026",
  title =        "The Rust Book - Unsafe Rust",
  url =          "https://doc.rust-lang.org/book/ch20-01-unsafe-rust.html",
  lastaccessed = "Janurary 20, 2026",
}

@misc{coq,
    author = "The Coq developers",
    node="",
    year="2026",
    url = "https://rocq-prover.org",
    title = "The Coq proof assistant",
    lastaccessed = "Janurary 20, 2026",
}

@misc{rvt,
    title = {"Rust Verification Tools"},
    author = {"The RVT developers"},
    year="2026",
    url = {https://project-oak.github.io/rust-verification-tools/about.html},
    lastaccessed = "Janurary 20, 2026",
}

@misc{framac,
    title = {"Frama-C"},
    author = {"Frama-C developers"},
    year="2026",
    url = {https://frama-c.com/},
    lastaccessed = "Janurary 20, 2026",
}

@misc{autoharness,
    title = {"Automatic Harness Generation"},
    author = {"Kani developers"},
    year="2026",
    url = {https://model-checking.github.io/kani/reference/experimental/autoharness.html},
    lastaccessed = "Janurary 20, 2026",
}

@misc{verify-rust-std,
    title = {"Verify Rust Standard Library Effort"},
    author = {"Kani developers"},
    year="2026",
    url = {https://model-checking.github.io/verify-rust-std},
    lastaccessed = "Janurary 20, 2026",
}

@misc{mut-in-out-pair,
  doi = {10.34727/2024/ISBN.978-3-85448-065-5_19},
  url = {https://repositum.tuwien.at/handle/20.500.12708/200786},
  author = {Lahiri, Shuvendu K.},
  language = {en},
  title = {Evaluating LLM-driven User-Intent Formalization for Verification-Aware Languages},
  publisher = {TU Wien},
  year = {2024},
  copyright = {Creative Commons Attribution 4.0 International}
}

@article{safe4u,
  author       = {Huan Li and
                  Bei Wang and
                  Xing Hu and
                  Xin Xia},
  title        = {Safe4U: Identifying Unsound Safe Encapsulations of Unsafe Calls in
                  Rust using LLMs},
  journal      = {Proc. {ACM} Softw. Eng.},
  volume       = {2},
  number       = {{ISSTA}},
  pages        = {457--480},
  year         = {2025},
  url          = {https://doi.org/10.1145/3728890},
  doi          = {10.1145/3728890},
  bibsource    = {dblp computer science bibliography, https://dblp.org}
}

@inproceedings{autospec,
  title={Enchanting program specification synthesis by large language models using static analysis and program verification},
  author={Wen, Cheng and Cao, Jialun and Su, Jie and Xu, Zhiwu and Qin, Shengchao and He, Mengda and Li, Haokun and Cheung, Shing-Chi and Tian, Cong},
  booktitle={International Conference on Computer Aided Verification},
  pages={302--328},
  year={2024},
  url={https://doi.org/10.1007/978-3-031-65630-9_16},
  organization={Springer}
}

@inproceedings{specgen,
  title={SpecGen: Automated Generation of Formal Program Specifications via Large Language Models},
  author={Ma, Lezhi and Liu, Shangqing and Li, Yi and Xie, Xiaofei and Bu, Lei},
  booktitle={2025 IEEE/ACM 47th International Conference on Software Engineering (ICSE)},
  pages={666--666},
  year={2025},
  url={https://dl.acm.org/doi/10.1109/ICSE55347.2025.00129},
  organization={IEEE Computer Society}
}

@article{rustbelt,
  title={RustBelt: Securing the foundations of the Rust programming language},
  author={Jung, Ralf and Jourdan, Jacques-Henri and Krebbers, Robbert and Dreyer, Derek},
  journal={Proceedings of the ACM on Programming Languages},
  volume={2},
  number={POPL},
  pages={1--34},
  year={2017},
  url={https://dl.acm.org/doi/10.1145/3158154},
  publisher={ACM New York, NY, USA}
}

@inproceedings{rusthornbelt,
  title={RustHornBelt: a semantic foundation for functional verification of Rust programs with unsafe code},
  author={Matsushita, Yusuke and Denis, Xavier and Jourdan, Jacques-Henri and Dreyer, Derek},
  booktitle={Proceedings of the 43rd ACM SIGPLAN International Conference on Programming Language Design and Implementation},
  pages={841--856},
  url={https://dl.acm.org/doi/10.1145/3519939.3523704},
  year={2022}
}

@article{refinedrust,
  title={Refinedrust: A type system for high-assurance verification of Rust programs},
  author={G{\"a}her, Lennard and Sammler, Michael and Jung, Ralf and Krebbers, Robbert and Dreyer, Derek},
  journal={Proceedings of the ACM on Programming Languages},
  volume={8},
  number={PLDI},
  pages={1115--1139},
  year={2024},
  doi={10.1145/3656422},
  publisher={ACM New York, NY, USA}
}

@article{gillianrust,
  title={A hybrid approach to semi-automated Rust verification},
  author={Ayoun, Sacha-{\'E}lie and Denis, Xavier and Maksimovi{\'c}, Petar and Gardner, Philippa},
  journal={Proceedings of the ACM on Programming Languages},
  volume={9},
  number={PLDI},
  pages={970--992},
  year={2025},
  doi={10.5281/zenodo.15183201},
  publisher={ACM New York, NY, USA}
}

@article{verus,
  title={Verus: Verifying rust programs using linear ghost types},
  author={Lattuada, Andrea and Hance, Travis and Cho, Chanhee and Brun, Matthias and Subasinghe, Isitha and Zhou, Yi and Howell, Jon and Parno, Bryan and Hawblitzel, Chris},
  journal={Proceedings of the ACM on Programming Languages},
  volume={7},
  number={OOPSLA1},
  pages={286--315},
  year={2023},
  url={https://dl.acm.org/doi/10.1145/3586037},
  publisher={ACM New York, NY, USA}
}

@misc{safe,
      title={Automated Proof Generation for Rust Code via Self-Evolution}, 
      author={Tianyu Chen and Shuai Lu and Shan Lu and Yeyun Gong and Chenyuan Yang and Xuheng Li and Md Rakib Hossain Misu and Hao Yu and Nan Duan and Peng Cheng and Fan Yang and Shuvendu K Lahiri and Tao Xie and Lidong Zhou},
      year={2026},
      eprint={2410.15756},
      archivePrefix={arXiv},
      primaryClass={cs.SE},
      url={https://arxiv.org/abs/2410.15756}, 
}

@inproceedings{kani,
author = {VanHattum, Alexa and Schwartz-Narbonne, Daniel and Chong, Nathan and Sampson, Adrian},
title = {Verifying dynamic trait objects in rust},
year = {2022},
isbn = {9781450392266},
publisher = {Association for Computing Machinery},
address = {New York, NY, USA},
url = {https://doi.org/10.1145/3510457.3513031},
doi = {10.1145/3510457.3513031},
booktitle = {Proceedings of the 44th International Conference on Software Engineering: Software Engineering in Practice},
pages = {321–330},
numpages = {10},
location = {Pittsburgh, Pennsylvania},
series = {ICSE-SEIP '22}
}

@inproceedings{unsafecop,
author="Wang, Minghua
and Xue, Jingling
and Huang, Lin
and Zi, Yuan
and Wei, Tao",
editor="Platzer, Andre
and Rozier, Kristin Yvonne
and Pradella, Matteo
and Rossi, Matteo",
title="UnsafeCop: Towards Memory Safety for Real-World Unsafe Rust Code with Practical Bounded Model Checking",
booktitle="Formal Methods",
year="2025",
publisher="Springer Nature Switzerland",
address="Cham",
pages="307--324",
isbn="978-3-031-71177-0",
url={https://link.springer.com/chapter/10.1007/978-3-031-71177-0_19},
}

@inproceedings{seahorn,
  title={SeaHorn: A framework for verifying C programs (competition contribution)},
  author={Gurfinkel, Arie and Kahsai, Temesghen and Navas, Jorge A},
  booktitle={International Conference on Tools and Algorithms for the Construction and Analysis of Systems},
  pages={447--450},
  year={2015},
  doi={10.1007/978-3-662-46681-0_41},
  organization={Springer}
}

@inproceedings{smack,
  title={SMACK: Decoupling source language details from verifier implementations},
  author={Rakamari{\'c}, Zvonimir and Emmi, Michael},
  booktitle={Computer Aided Verification: 26th International Conference, CAV 2014, Held as Part of the Vienna Summer of Logic, VSL 2014, Vienna, Austria, July 18-22, 2014. Proceedings 26},
  pages={106--113},
  year={2014},
  doi={10.1007/978-3-319-08867-9_7},
  organization={Springer}
}

@inproceedings{panic-checker-klee,
author = {Zhang, Ying and Li, Peng and Ding, Yu and Wang, Lingxiang and Williams, Dan and Meng, Na},
title = {Broadly Enabling KLEE to Effortlessly Find Unrecoverable Errors in Rust},
year = {2024},
isbn = {9798400705014},
publisher = {Association for Computing Machinery},
address = {New York, NY, USA},
url = {https://doi.org/10.1145/3639477.3639714},
doi = {10.1145/3639477.3639714},
booktitle = {Proceedings of the 46th International Conference on Software Engineering: Software Engineering in Practice},
pages = {441–451},
numpages = {11},
location = {Lisbon, Portugal},
series = {ICSE-SEIP '24}
}

@misc{crux,
      title={Crux, a Precise Verifier for Rust and Other Languages}, 
      author={Stuart Pernsteiner and Iavor S. Diatchki and Robert Dockins and Mike Dodds and Joe Hendrix and Tristan Ravich and Patrick Redmond and Ryan Scott and Aaron Tomb},
      year={2024},
      eprint={2410.18280},
      archivePrefix={arXiv},
      primaryClass={cs.PL},
      url={https://arxiv.org/abs/2410.18280}, 
}

@InProceedings{pei-llm-reason-loop-inv,
  title = 	 {Can Large Language Models Reason about Program Invariants?},
  author =       {Pei, Kexin and Bieber, David and Shi, Kensen and Sutton, Charles and Yin, Pengcheng},
  booktitle = 	 {Proceedings of the 40th International Conference on Machine Learning},
  pages = 	 {27496--27520},
  year = 	 {2023},
  editor = 	 {Krause, Andreas and Brunskill, Emma and Cho, Kyunghyun and Engelhardt, Barbara and Sabato, Sivan and Scarlett, Jonathan},
  volume = 	 {202},
  series = 	 {Proceedings of Machine Learning Research},
  month = 	 {23--29 Jul},
  publisher =    {PMLR},
  url = 	 {https://proceedings.mlr.press/v202/pei23a.html}
}

@inproceedings{rank-llm-gen-loop-inv,
  title={Ranking llm-generated loop invariants for program verification},
  author={Chakraborty, Saikat and Lahiri, Shuvendu and Fakhoury, Sarah and Lal, Akash and Musuvathi, Madanlal and Rastogi, Aseem and Senthilnathan, Aditya and Sharma, Rahul and Swamy, Nikhil},
  booktitle={Findings of the Association for Computational Linguistics: EMNLP 2023},
  pages={9164--9175},
  doi={10.18653/v1/2023.findings-emnlp.614},
  year={2023}
}

@article{autoverus,
  title={Autoverus: Automated proof generation for rust code},
  author={Yang, Chenyuan and Li, Xuheng and Misu, Md Rakib Hossain and Yao, Jianan and Cui, Weidong and Gong, Yeyun and Hawblitzel, Chris and Lahiri, Shuvendu and Lorch, Jacob R and Lu, Shuai and others},
  journal={Proceedings of the ACM on Programming Languages},
  volume={9},
  number={OOPSLA2},
  pages={3454--3482},
  year={2025},
  url={https://dl.acm.org/doi/abs/10.1145/3763174},
  publisher={ACM New York, NY, USA}
}

@misc{alphaverus,
      title={AlphaVerus: Bootstrapping Formally Verified Code Generation through Self-Improving Translation and Treefinement}, 
      author={Pranjal Aggarwal and Bryan Parno and Sean Welleck},
      year={2024},
      eprint={2412.06176},
      archivePrefix={arXiv},
      primaryClass={cs.LG},
      url={https://arxiv.org/abs/2412.06176}, 
}

@article{llm_dafny,
  title={Towards ai-assisted synthesis of verified dafny methods},
  author={Misu, Md Rakib Hossain and Lopes, Cristina V and Ma, Iris and Noble, James},
  journal={Proceedings of the ACM on Software Engineering},
  volume={1},
  number={FSE},
  pages={812--835},
  year={2024},
  url={https://dl.acm.org/doi/10.1145/3643763},
  publisher={ACM New York, NY, USA}
}

@inproceedings{dafny,
  title={Dafny: An automatic program verifier for functional correctness},
  author={Leino, K Rustan M},
  booktitle={International conference on logic for programming artificial intelligence and reasoning},
  pages={348--370},
  year={2010},
  url={https://link.springer.com/chapter/10.1007/978-3-642-17511-4_20},
  organization={Springer}
}

@article{boogie,
  title={This is boogie 2},
  author={Leino, K Rustan M},
  journal={manuscript KRML},
  volume={178},
  number={131},
  pages={9},
  year={2008},
  url={https://www.microsoft.com/en-us/research/wp-content/uploads/2016/12/krml178.pdf},
  publisher={Citeseer}
}

@inproceedings{he2025ranking,
  title={Ranking Formal Specifications using LLMs},
  author={He, Mike and Ang, Zhendong and Desai, Ankush and Gupta, Aarti},
  booktitle={Proceedings of the 1st ACM SIGPLAN International Workshop on Language Models and Programming Languages},
  pages={51--56},
  doi={10.1145/3759425.3763386},
  year={2025}
}

@inproceedings{pmodel,
  author       = {Ankush Desai and
                  Vivek Gupta and
                  Ethan K. Jackson and
                  Shaz Qadeer and
                  Sriram K. Rajamani and
                  Damien Zufferey},
  editor       = {Hans{-}Juergen Boehm and
                  Cormac Flanagan},
  title        = {{P:} safe asynchronous event-driven programming},
  booktitle    = {{ACM} {SIGPLAN} Conference on Programming Language Design and Implementation,
                  {PLDI} '13, Seattle, WA, USA, June 16-19, 2013},
  pages        = {321--332},
  publisher    = {{ACM}},
  year         = {2013},
  url          = {https://doi.org/10.1145/2491956.2462184},
  doi          = {10.1145/2491956.2462184},
  bibsource    = {dblp computer science bibliography, https://dblp.org}
}

@inproceedings{req_ambiguity_detect,
  title={Requirements ambiguity detection and explanation with llms: An industrial study},
  author={Bashir, Sarmad and Ferrari, Alessio and Khan, Abbas and Strandberg, Per Erik and Haider, Zulqarnain and Saadatmand, Mehrdad and Bohlin, Markus},
  booktitle={2025 IEEE International Conference on Software Maintenance and Evolution (ICSME)},
  pages={620--631},
  year={2025},
  doi={10.1109/ICSME64153.2025.00063},
  organization={IEEE}
}

@misc{llm_for_req_eng,
      title={Large Language Models (LLMs) for Requirements Engineering (RE): A Systematic Literature Review}, 
      author={Mohammad Amin Zadenoori and Jacek Dąbrowski and Waad Alhoshan and Liping Zhao and Alessio Ferrari},
      year={2025},
      eprint={2509.11446},
      archivePrefix={arXiv},
      primaryClass={cs.SE},
      url={https://arxiv.org/abs/2509.11446}, 
}

@article{auto_software_doc,
  title={Automating software documentation: Employing llms for precise use case description},
  author={Naimi, Lahbib and Jakimi, Abdeslam and Saadane, Rachid and Chehri, Abdellah and others},
  journal={Procedia Computer Science},
  volume={246},
  pages={1346--1354},
  year={2024},
  doi={10.1016/j.procs.2024.09.568},
  publisher={Elsevier}
}

@article{cohen_kappa,
  author = {Cohen, Jacob},
  title = {A Coefficient of Agreement for Nominal Scales},
  journal = {Educational and Psychological Measurement},
  volume = {20},
  number = {1},
  pages = {37-46},
  doi={10.1177/001316446002000104},
  year = {1960}
}

@misc{kapilot_artifact,
    title = {KaPilot Artifact},
    author = {Minghua Wang and Yuxi Ling and Mingzhi Gao and Yuwei Liu and Lin Huang},
    year={2026},
    doi= {10.6084/m9.figshare.32834699.v3},
    lastaccessed = "July 17, 2026",
}

\end{document}